\documentclass[%
reprint,
amsmath,amssymb,
aps,
onecolumn,
]{revtex4-2}\usepackage{parskip} 
\usepackage{tikz} 
\usepackage{amsmath}
\usepackage{hyperref}
\hypersetup{colorlinks}
\usepackage{graphicx}
\usepackage{subcaption}
\usepackage{amsfonts}
\usepackage{bbold}
\usepackage{braket}
\usepackage{ulem}
\usepackage{cancel}
\usepackage{multirow} 

\newlength{\figwidthThird}
\newlength{\figwidthHalf}
\begin{document}
\title{Hamiltonian Switching Control of Noisy Bipartite Qubit Systems}

\author{Zhibo Yang}
\affiliation{Department of Chemistry, University of California, Berkeley, California 94720, USA}
\affiliation{Berkeley Center for Quantum Information and Computation, Berkeley, California 94720, USA}
\author{Robert L. Kosut}
\affiliation{SC Solutions, Sunnyvale, California 94085 USA}
\affiliation{Department of Chemistry,
  Princeton University, Princeton, NJ, 08544, USA}
\author{K. Birgitta Whaley}
\affiliation{Department of Chemistry, University of California, Berkeley, California 94720, USA}
\affiliation{Berkeley Center for Quantum Information and Computation, Berkeley, California 94720, USA}
\begin{abstract}
    We develop a Hamiltonian switching ansatz for bipartite control that is inspired by the Quantum Approximate Optimization Algorithm (QAOA), to mitigate environmental noise on qubits. We illustrate the approach with application to the protection of quantum gates performed on i) a central spin qubit coupling to bath spins through isotropic Heisenberg interactions, ii) superconducting transmon qubits coupling to environmental two-level-systems (TLS) through dipole-dipole interactions, and iii) qubits coupled to both TLS and a Lindblad bath. The control field is classical and acts only on the system qubits. We use reinforcement learning with policy gradient (PG) to optimize the Hamiltonian switching control protocols, using a fidelity objective defined with respect to specific target quantum gates. 
    We use this approach to demonstrate effective suppression of both coherent and dissipative noise, with numerical studies achieving target gate implementations with fidelities over 0.9999 (four nines) in the majority of our test cases and showing improvement beyond this to values of 0.999999999 (nine nines) upon a subsequent optimization by Gradient Ascent Pulse Engineering (GRAPE). We analyze how the control depth, total evolution time, number of environmental TLS, and choice of optimization method affect the fidelity achieved by the optimal protocols and reveal some critical behaviors of bipartite control of quantum gates. 
\end{abstract}

\date{\today}
\maketitle

\section{Introduction}
Quantum computation has a number of promising potential applications. However, currently available noisy intermediate scale quantum (NISQ) devices suffer from various sources of noise, which limit their performance \cite{Preskill2018quantumcomputingin,Koch2016}. Error mitigation methods are needed to improve the fidelity of quantum circuits in order to obtain meaningful results from the computations. To generate unitary controls that mitigate against qubit noise due to unwanted interactions with external qubit or spin systems, we introduce here a Hamiltonian switching technique inspired by the Quantum Approximate Optimization Algorithm (QAOA). First studied in the context of the combinatorial optimization problems \cite{Farhi2014}, QAOA employs a set of piece-wise constant Hamiltonians to parameterize quantum circuits with a relatively small set of parameters. The QAOA methodology has also received interest as an ansatz for quantum control and error mitigation. Several recent works have investigated robust control under classical noise using QAOA with reinforcement learning \cite{Bukov2018, Niu2019}, sequential convex programming (SCP) \cite{KosutGB:13,Dong2019} and policy gradient (PG) \cite{Yao2020, Yao2020Robust, Yao2020cd}. These methods were shown to outperform prevalent gradient-based control algorithms for the majority of test-cases on both a single qubit and a coupled spin chain having classical Hamiltonian uncertainties, suggesting an intrinsic robustness of QAOA-based controls.

A typical control Hamiltonian $H_c(t)$ is dependent on a set of control variables $\{v_j(t),j=1\dots,m\}$, where often $H_c(t)$ is linear in these variables: $H_c(t)=\sum_{j=1}^mv_j(t)H_j$. In our Hamiltonian switching ansatz, the operational time of the gate is divided into $m$ variable length intervals that we refer to as "hold times" and $H_c(t)$ is held constant during each hold time at one of the $m$ Hamiltonians in a predefined set, $\{H_j,j=1\ldots,m\}$. During each hold time only one control is active and is set to its maximum value; the remaining controls are set to zero.

In this work we apply the Hamiltonian switching ansatz to control of the \textit{central spin} (CS) model. Sometimes referred to as the \textit{spin star} model, the CS model is relevant to various types of solid-state quantum devices \cite{Schliemann_2003,Breuer2004,Arenz2014}. Extensive work has been done on different aspects of CS models for quantum information processing, such as reduced spin dynamics \cite{Breuer2004, Fischer2007, Bhattacharya2017,Jing2018}, coherence times\cite{Shenvi2005,Shenvi2005a,deSousa2005} and bipartite control \cite{Grace2007,Grace2007a,Grace2010,Arenz2014,Kosut2019}. In the CS model, the central spin, which can be used as a qubit for quantum computation, is coupled to a collection of bath spins through isotropic \cite{Gaudin1976} or dipole-dipole \cite{Hutton2004} Heisenberg spin-spin interactions. The CS model is also relevant to coherent noise in superconducting transmon qubits. In these devices, multiple \textit{two-level-systems} (TLS) couple coherently to the qubit through dipole-dipole Heisenberg couplings due to the defects introduced by device fabrication \cite{Muller2019}. The Hamiltonian then takes the same form as a CS model with dipole-dipole couplings. The transmon qubit has been one of the most promising candidates for near-term applications and has been notably used for Google's quantum supremacy experiment \cite{Arute2019}. The energy splittings, coupling strengths, and mechanism of TLS coupling have been experimentally characterized in several works \cite{Simmonds2004, Neeley2008, Cole2010}. Bipartite control of a single transmon qudit coupled to both a TLS bath and a Lindblad bath, which provides a more realistic model for real physical devices, has also been studied \cite{Reich_2015}.

Despite this activity, relatively little work has been done on controlling central spin systems to implement target gates with sufficiently high fidelity for fault-tolerant quantum computation. Specifically, it is highly desirable to implement elementary quantum gates with fidelity higher than the thresholds of practically relevant quantum error correction codes. Since the error thresholds for non-local concatenation codes and practical surface codes are of order $10^{-4}$ \cite{Terhal2015,Aliferis2007,Campbell2017}, a gate fidelity higher than  $1-10^{-4} = 0.9999$ (we refer to this as "four nines") is a reasonable target for successful control of quantum computers. In this work, we will test the performance of bipartite Hamiltonian switching control on CS systems under the effect of coupling to both a spin bath (TLS bath) and a Lindblad bath, for the implementation of high-quality quantum logic gates with fidelity higher than $1-10^{-4}$. The PG algorithms will be used for the classical optimization of the control ansatz.

Our results show that the bipartite Hamiltonian switching control can readily implement single qubit gates on a qubit isotropically coupled to bath spins with fidelity over six nines, and single and two qubit gates on qubits coupled to TLS through dipole-dipole couplings with fidelity over four nines. The fidelity of single qubit gates with dipole-dipole coupled TLS can be further improved to nine nines by adding a secondary optimization using GRadient Ascent Pulse Engineering (GRAPE  \cite{Khaneja2005}). For systems with a secondary Lindblad bath, the fidelity achieved lies generally between two to four nines. We also reveal the trends of fidelity dependence on total time, control depth and the number of bath spins/TLS. The fidelity of different gates on the same system have similar fidelity level and control properties. 

The remainder of the paper is structured as follows. We will first introduce our physical models in Sec.~\ref{sec:PhysicalModels}, followed by presentation of the bipartite Hamiltonian switching control ansatz and secondary optimization by GRAPE in Sec.~\ref{sec:controls}. We then present the fidelity measure and the classical algorithms used for optimization in Sec.~\ref{sec:Methods}. Numerical results and analysis are presented in Sec.~\ref{sec:results} and the features of the results discussed in Sec. \ref{sec:discussion}. We then discuss the fidelity differences for different test cases and present possible improvements in Sec. \ref{sec:discussion}. Sec. \ref{sec:conclusions} concludes with a summary and outlook.

\section{Physical Models} \label{sec:PhysicalModels}
    We set $\hbar =1$ throughout this work. $\sigma^i_q$, $i\in \{z,y,z,+,-\}$ are Pauli operators. The number of bath spins/TLS is $n$. We study two types of central spin (CS) models with different types of practically relevant couplings to the bath spins. We also consider additional Lindblad decoherence due to a secondary bath acting on both the qubits and the primary bath of spins.\\
    A general Hamiltonian for the controlled dynamics is 
    \begin{equation}
        H(t) = H_S + H_c(t) + H_{\text{env}} + H_I,
    \end{equation}
    where $H_S$ and $H_{\text{env}}$ describes the intrinsic dynamics of the system (qubits) and primary bath (bath spins/ TLS) respectively, and $H_I$ describes the coupling between system and primary spin bath. These terms are all time-independent. $H_c(t)$ describes the time-dependent control applied to the system qubits, which will be discussed in detail in the next section.\\
    We study the control problem on single and two-qubit systems. The system Hamiltonian $H_S$ for the one-qubit system is just its energy splitting
    \begin{equation}
        H_S^{\text{1qubit}} = -\dfrac{E}{2} \sigma^z_0.
    \end{equation}
    For the 2-qubit system, $H_S$ contains an extra coupling term between the two qubits to enable implementation of 2-qubit gates:
    \begin{equation}
        H_S^{\text{2qubit}} = -\dfrac{E_0}{2} \sigma^z_0 -\dfrac{E_1}{2} \sigma^z_1 + \gamma \sigma^z_0 \sigma^z_1.
    \end{equation}
    Note that neither $H_S^{\text{1qubit}}$ nor $H_S^{\text{2qubit}}$ allow implementation of arbitrary 1-qubit or 2-qubit quantum gates, respectively. This capability will be added by the control Hamiltonians $H_c(t)$. Note that the enumeration of qubits is different in the 1-qubit and 2-qubit systems. For the 1-qubit system, the system qubit is indexed as 0 and the bath spins start from index 1. For the 2-qubit system, the system qubits are indexed as 0 and 1 and the bath spin index starts at 2.\\
    
    \subsection{Isotropic Couplings}
    In the model with an isotropic Heisenberg interaction, all of the environmental spins couple to the central spin through a $\sigma^x \sigma^x + \sigma^y \sigma^y + \sigma^z \sigma^z$ type of interaction. The coupling Hamiltonian is then
    \begin{equation} \label{eq:CSnoCtr}
        H_I^{\text{iso}} = \sum_{q=1}^{n} A_q \sum_{s=x,y,z} \sigma^s_0 \sigma^s_q.
    \end{equation}
    Using the parameters of \cite{Arenz2014}, we set the energy splitting of the system qubit to be $E=1$. The coupling strengths are set to be $A_q = 1$ when they are all equal, and to be uniformly distributed between 1 and 2 ($A_q \sim \mathcal{U}[1, 2]$) when they are not equal. The energy unit is chosen to be arbitrary for simplicity and generality. Note that the coupling strength chosen for this model is very strong compared to that used in the more physically-relevant model in the next subsection. 
    
    \subsection{Dipole-dipole Couplings} \label{sec:dp_model}
    In this model, environmental TLS are coupled to the system through a $\sigma^+\sigma^- + \sigma^-\sigma^+ \equiv 1/2(\sigma^x\sigma^x + \sigma^y\sigma^y)$ type dipole-dipole Heisenberg interaction. This model is relevant to decoherence in superconducting transmon qubits, one important component of which is caused by the qubit coupling to TLS that are associated with device defects \cite{Muller2019}. Among several possible mechanisms for qubit-TLS coupling, we focus on one that is induced by charge fluctuations, resulting in a dipole-dipole coupling. The interaction Hamiltonian takes the form   
    \begin{equation}\label{eq:TLSint}
        H_I^{\text{dipole}} = \sum_{q=1}^{n} \frac{A_q}{2} (\sigma^+_0\sigma^-_q + \sigma^-_0\sigma^+_q ) \equiv \sum_{q=1}^{n} \frac{A_q}{4} (\sigma^x_0\sigma^x_q + \sigma^y_0\sigma^y_q )
    \end{equation}
    and the TLS energy splittings are
    \begin{equation}
        H_{\text{env}}^{\text{dipole}} = -\sum_{q=1}^n \frac{\Delta_q}{2}\sigma^z_q.
    \end{equation}
    For calculations with this model we use a set of experimentally relevant parameters. The TLS energy splittings $\Delta_q$ have a similar magnitude as the qubit energy splitting $E$ (i.e., $\Delta_q \approx E$), and the coupling strength $A_q$ is normally 2 to 3 orders of magnitude smaller \cite{Simmonds2004,Neeley2008,Cole2010}. The coupling constants in real devices are usually not equal, so we focus here on the variable coupling case. We use $E = 8 \, \text{GHz}$, $\Delta_q=8+0.8q\, \text{GHz}$, and $A_q$ having a variable value uniformly distributed in the range $4 - 40 \, \text{MHz}$ \cite{Simmonds2004,Neeley2008,Cole2010}. The system energy splitting $E$ is equivalent to an angular frequency $\omega = 8\times 2\pi \cdot 10^9 \, \text{rad} \cdot s^{-1}$. To compare with the unit-less convention used in the isotropic model in the previous subsection, this angular frequency is normalized to be 1 and consequently the unit time in our simulation is $\frac{1}{16\pi}$ ns.
    
    In the two-qubit system with dipolar coupling to TLS, the bath and interaction Hamiltonians take the form
    \begin{equation}\label{eq:dp2qb}
        \begin{aligned}
        H_{\text{env}}^{\text{dipole-2qubit}} &= -\sum_{i=0}^1 \sum_{{q_i}=1}^{n_i} \frac{\Delta_{q_i}}{2}\sigma^z_{q_i} \\
        H_I^{\text{dipole-2qubit}} &= \sum_{i=0}^1 \sum_{{q_i}=1}^{n_i} \frac{A_{q_i}}{2} (\sigma^+_{i}\sigma^-_{q_i} + \sigma^-_{i} \sigma^+_{q_i}) \equiv \sum_{i=0}^1 \sum_{{q_i}=1}^{n_i} \frac{A_{q_i}}{4} (\sigma^x_{i}\sigma^x_{q_i} + \sigma^y_{i} \sigma^y_{q_i}).
        \end{aligned}
    \end{equation}
    Here the index $i$ enumerates the system qubits and $n_i$ is the number of TLS coupling to the system qubit $i$. We set $E_0 = 8.0$ GHz and $E_1 = 8.4$ GHz. All other parameter strengths are the same as for the single qubit. 
    
    Since the classical simulation cost scales exponentially with system size, the largest system simulated here is of six bath spins/TLS. However, as shown in Sec. \ref{sec:reduced_dyna}, within the time-frame of our simulations and in the absence of control, several bath TLS can effectively represent a larger and even infinite spin bath. Fig. \ref{fig:dp_dyna} shows that the dynamics of a single qubit coupled to a different finite number of TLS or an infinite TLS bath have very good agreement up to 2 ns, which is beyond the maximal time of most simulations in this work.
    \subsection{Reduced Dynamics of Different Bath Sizes} \label{sec:reduced_dyna}
    The dynamics of central spin systems with specific system Hamiltonians have been solved analytically in \cite{Breuer2004} for dipole-dipole couplings and in \cite{Fischer2007} for isotropic couplings. These solutions cannot be directly applied to our simulations as they require conservation of the $z$-component of total angular momentum, which is not the case in our control ansatz. However, we can utilize these solutions in the absence of control to compare the dynamics of our simulated smaller systems with larger systems and even an infinite spin bath. 
    
\begin{figure}[h!]
	 	\centering
	 	\begin{subfigure}[b]{\figwidthHalf}
	 	     \includegraphics[width=\linewidth]{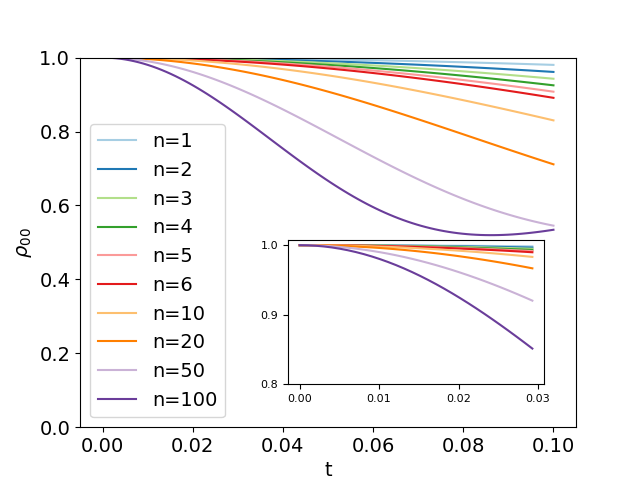}
	 	     \caption{}
	 	     \label{fig:Heis_dyna}
	 	\end{subfigure}
	 	\begin{subfigure}[b]{\figwidthHalf}
	 	     \includegraphics[width=\linewidth]{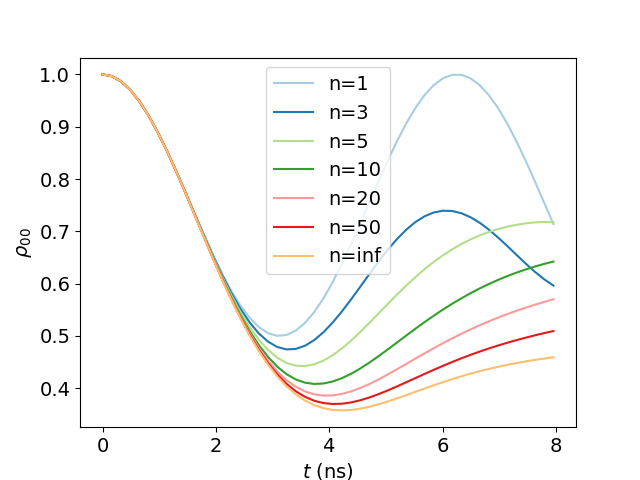}
	 	     \caption{}
	 	     \label{fig:dp_dyna}
	 	\end{subfigure}
	 	\caption{The $\ket{0}$ state population of a qubit coupled to different numbers of bath spins through (a) isotropic coupling with strength 1 (arbitrary unit) and (b) dipole-dipole coupling of strength 40 MHz , assuming the qubit energy splitting to be 1 (arbitrary unit) and 8 GHz respectively. The aforementioned coupling strengths are of the same scale as our simulated systems. The initial state of the qubit is $\ket{0}$ and the initial state of the bath is an infinite temperature mixed state.}
 		\label{fig:cs_dyna}
	\end{figure}

    The dynamics of central spin systems of different sizes with no control fields are plotted in Fig. \ref{fig:cs_dyna}. The initial state of the qubit is $\ket{0}$, and the initial state of the bath is an infinite temperature mixed state for both plots. \\
	
    Panel (a) shows that the dynamics of systems with less than 10 bath spins agree well only within $t=0.02$. However, as will be presented in the next section, the typical timescale of a working control protocol for the isotropic coupling is well above 1, which means that the simulated central spin systems cannot represent a larger bath. \\
	
	Panel (b) shows that, for a superconducting qubit coupled to TLS through dipole-dipole couplings, the dynamics agree well up to 2 ns for different number of TLS, even for a infinite number of TLS. In comparison, the protocols in this work typically take less than 0.2 ns, which is well below this timescale. Therefore, the optimized protocols for smaller dipolar TLS coupled systems should be applicable to a larger or even infinite TLS bath.
    \subsection{Secondary Bath}
    We refer to the finite number of environmental spins as the \textit{primary bath}. To better simulate real physical systems, in some calculations we further also consider a \textit{secondary bath}, described by Lindblad dynamics, that acts on both the system qubits and the environmental spins (TLS). For this we choose a model modified from that in \cite{Reich_2015}, which is a superconducting device coupled to TLS with dipole-dipole coupling under the effect of spontaneous emission ($T_1$ decay). This model is an extension to our model in Sec. \ref{sec:dp_model}. The overall dynamics for this model can be described by
    \begin{equation}\label{eq:Lind}
        \frac{d \rho}{d t}=-i\left[H(t), \rho \right]+\sum_{k} \left(L_{k} \rho L_{k}^{\dagger}-\frac{1}{2}\left\{L_{k}^{\dagger} L_{k}, \rho\right\}\right).
    \end{equation}
    Here the Lindblad operators for spontaneous emission are $L_0 = 1/\sqrt{T_1^S} \sigma_0^-$ for the system and $L_q = 1/\sqrt{T_1^{\text{dipole}}} \sigma_q^-$ for each of the TLS, where $T_1^S$ and $T_1^{\text{dipole}}$ are the $T_1$ time of the qubit and TLS respectively.

\section{Bipartite Control with Hamiltonian Switching} \label{sec:controls}
    The idea of QAOA is to switch between two time-constant Hamiltonians according to an ansatz designed for a specific computational task \cite{Farhi2014}. The QAOA approach is naturally related to bang-bang control when used as a method for control pulse parameterization. In the current implementation of the Hamiltonian switching method, we also employ two control Hamiltonians $H_A$ and $H_B$. The parameters are then the time duration over which each piece-wise constant Hamiltonian is applied. In the $i$-th round out of $p$ total rounds, we denote the time applying $H_A$ to be $\alpha_i$ and the time applying $H_B$ to be $\beta_i$. $p$ is called the \textit{control depth},i.e. the number of switching times is $2p$ and it is the total number of parameters. The control sequence $\boldsymbol{\theta} \equiv \{ \alpha_i, \beta_i\}_{i=1}^p$ is then optimized classically. The unitary evolution of the full system can then be written as 
	\begin{equation}\label{eq:UniQAOA}
	U(\boldsymbol{\theta}) = e^{-iH_B\beta_p}e^{-iH_A\alpha_p} \ldots e^{-iH_B\beta_1}e^{-iH_A\alpha_1},
	\end{equation}
	with the \textit{total duration} of the controls given by $T = \sum_{i=1}^{p} (\alpha_i + \beta_i)$. We define a fidelity measure $F(\boldsymbol{\theta})$ in the next section that serves as the performance merit of the control. We then use classical optimization methods to find the optimal control sequence $\boldsymbol{\theta}^*$, defined as
	\begin{equation}\label{key}
	\boldsymbol{\theta}^* \equiv \left\{\alpha_{i}^{*}, \beta_{i}^{*}\right\}_{i=1}^{p}=\underset{\left\{\alpha_{i}, \beta_{i}\right\}_{i=1}^{p}, \sum_{i=1}^{p} (\alpha_i + \beta_i) = T}{\arg \max } F\left(\left\{\alpha_{i}, \beta_{i}\right\}_{i=1}^{p}\right).
	\end{equation}
	The classical optimizer employed in this work is reinforcement learning with policy gradient (PG), which will be introduced in Sec. \ref{sec:OptAlg}.

    \subsection{Control Ansatz}\label{sec:ansatz}
    The goal of the Hamiltonian switching control in this work is to implement arbitrary quantum logic gates on the system qubits, which requires that the control ansatz be universal on the logical subspace. Thus, the alternating terms of the Hamiltonian switching ansatz should be able to generate arbitrary unitary transformations on the system Hilbert space in the absence of coupling to the environment. 
    \subsubsection{Single-qubit}
    For single-qubit control, the ansatz we choose is derived from an experimentally relevant form of the Landau–Zener Hamiltonian \cite{Bason2012, Hegerfeldt2013}. We apply a piece-wise constant control pulse with constant strength and alternating direction in the form of $\sigma^x$ on the system qubit, with an always-on system $\sigma^z$ term representing the system energy splitting \cite{Bukov2018}. The ansatz is then expanded by turning on coupling to environmental degrees of freedom as described in the previous section. The control is only on the system part, which makes the control problem bipartite. 
    
    For isotropic coupling, we then evolve the system under the following two piece-wise constant Hamiltonians:
	\begin{equation}
        \begin{aligned}
	   H_A^{\text{iso}} = -\dfrac{E}{2} \sigma^z_0 + 2E\sigma_0^x  + \sum_{q=1}^{n} A_q \sum_{s=1}^{3} \sigma^s_0 \sigma^s_q =  H_S^{\text{1qubit}} + 2E\sigma_0^x +H_I^{\text{iso}}\\
	   H_B^{\text{iso}} = -\dfrac{E}{2} \sigma^z_0 - 2E\sigma_0^x  + \sum_{q=1}^{n} A_q \sum_{s=1}^{3} \sigma^s_0 \sigma^s_q = H_S^{\text{1qubit}} - 2E\sigma_0^x +H_I^{\text{iso}}.
	\end{aligned}
	\end{equation}
	
	For the system with dipolar coupling to TLS, the corresponding piece-wise constant Hamiltonians are 
	\begin{equation} \label{eq:TLS_AB}
	\begin{aligned}
	    H_A^{\text{dipole}} = -\frac{E}{2} \sigma^z_0 + 2E \sigma_0^x -\sum_{q=1}^n \frac{\Delta_q}{2}\sigma^z_q + \sum_{q=1}^{n} \frac{A_q}{2} (\sigma^+_0\sigma^-_q + \sigma^-_0\sigma^+_q )= H_S^{\text{1qubit}} + 2E\sigma_0^x + H_I^{\text{dipole}} +H_{\text{env}}^{\text{dipole}}\\
	    H_B^{\text{dipole}} = -\frac{E}{2} \sigma^z_0 - 2E \sigma_0^x -\sum_{q=1}^n \frac{\Delta_q}{2}\sigma^z_q + \sum_{q=1}^{n} \frac{A_q}{2} (\sigma^+_0\sigma^-_q + \sigma^-_0\sigma^+_q )= H_S^{\text{1qubit}} - 2E\sigma_0^x +H_I^{\text{dipole}} +H_{\text{env}}^{\text{dipole}}.
	\end{aligned}
	\end{equation}

        To further improve the fidelity for this model, we considered application of a 4-Hamiltonian switching scheme by adding a $\sigma^y$ control term, as detailed in Appendix \ref{sec:4Ham}. The combination of alternating signs of $\sigma^y$ and $\sigma^y$ control terms then gives 4 switching Hamiltonians. We also further tested a secondary optimization with GRadient Ascent Pulse Engineering (GRAPE) \cite{Khaneja2005} on the optimization results of the 2-Hamiltonian switching control. With this additional secondary optimization, the unitary evolution operator becomes
        \begin{equation}\label{eq:UniGRAPE}
	U(\mathbf{c},\boldsymbol{\theta}) = \prod_{i=1}^p \exp(-i(H_d^{\text{dipole}} +c_{b_i} H_c)\beta_i) \exp(-i(H_d^{\text{dipole}} +c_{a_i} H_c)\alpha_i),
	\end{equation}
        where $H_d^{\text{dipole}} = H_S^{\text{1qubit}} +H_I^{\text{dipole}} +H_{\text{env}}^{\text{dipole}}$ and $H_c = 2E\sigma_0^x$, the time-duration protocol $\boldsymbol{\theta} \equiv \{ \alpha_i, \beta_i\}_{i=1}^p$ is the optimized result for Hamiltonian switching, and the piecewise-constant pulse-strength protocol $\mathbf{c} \equiv \{ c_{a_i}, c_{b_i}\}_{i=1}^p$ is optimized using GRAPE. The strengths $\mathbf{c}$ are initialized as alternating $\pm 1$, which are the strengths in the Hamiltonian switching protocol.
        
	We also tested a model that is a single-qubit analog of the 4-level qudit model in \cite{Reich_2015} by changing the control term to $\pm E\sigma^z_0$ in Eq. \ref{eq:TLS_AB}, resulting in the following switching Hamiltonians
        \begin{equation} \label{eq:non_uni_ansatz}
	\begin{aligned}
	    H_A^{\text{NonUni}} = -\frac{E}{2} \sigma^z_0 + 2E \sigma_0^z -\sum_{q=1}^n \frac{\Delta_q}{2}\sigma^z_q + \sum_{q=1}^{n} \frac{A_q}{2} (\sigma^+_0\sigma^-_q + \sigma^-_0\sigma^+_q )= H_S^{\text{1qubit}} + 2E\sigma_0^z + H_I^{\text{dipole}} +H_{\text{env}}^{\text{dipole}}\\
	    H_B^{\text{NonUni}} = -\frac{E}{2} \sigma^z_0 - 2E \sigma_0^z -\sum_{q=1}^n \frac{\Delta_q}{2}\sigma^z_q + \sum_{q=1}^{n} \frac{A_q}{2} (\sigma^+_0\sigma^-_q + \sigma^-_0\sigma^+_q )= H_S^{\text{1qubit}} - 2E\sigma_0^z +H_I^{\text{dipole}} +H_{\text{env}}^{\text{dipole}}.
	\end{aligned}
	\end{equation}
	Without the TLS bath this Hamiltonian can only generate $Z$-rotations  and is thus not universal in SU(2). However, as noted in \cite{Reich_2015}, the dipole-dipole coupling between the 4-level qudit and TLS introduces transversal terms to the Hamiltonian and thereby changes the algebraic structure of this ansatz, enabling it to generate transformations in a larger subspace of the corresponding unitary group. It is expected that the same argument might also hold for the model of Eq. \ref{eq:non_uni_ansatz}, since the Lie algebra generated by the $\sigma^z$ terms in Eq. \ref{eq:non_uni_ansatz} and the dipolar coupling terms has higher rank than the trivial Lie algebra generated by the drift and control term in Eq. \ref{eq:non_uni_ansatz} alone. Therefore, the effect of Hamiltonian switching with this non-universal ansatz Eq. \ref{eq:non_uni_ansatz} will also be tested.

	\subsubsection{Two-qubits}
	For a universal ansatz on the SU(4) group for two-qubit gates, the control ansatz we use is based on that proposed in \cite{Lloyd2018quantum}. The idea is to alternate 2-local $\sigma^z \sigma^z$ terms and single-qubit $\sigma^x$ terms. For a control ansatz, we treat the $\sigma^z$ terms as the drift part and alternate the sign of $\sigma^x$ terms, which will not change the generated Lie algebra. This results in the following QAOA Hamiltonians
	\begin{equation}\label{eq:2qb_ansatz}
    \begin{aligned}
    H_A^{\text{dipole-2qubit}} &= -\dfrac{E_0}{2} \sigma^z_0 -\dfrac{E_1}{2} \sigma^z_1 + \gamma \sigma^z_0 \sigma^z_1 + E_0(\sigma^x_0+\sigma^x_1) +H_{\text{env}}^{\text{dipole-2qubit}} + H_I^{\text{dipole-2qubit}}\\
    H_B^{\text{dipole-2qubit}} &= -\dfrac{E_0}{2} \sigma^z_0 -\dfrac{E_1}{2} \sigma^z_1 + \gamma \sigma^z_0 \sigma^z_1 - E_0(\sigma^x_0+\sigma^x_1) +H_{\text{env}}^{\text{dipole-2qubit}} + H_I^{\text{dipole-2qubit}},
    \end{aligned}
    \end{equation}
    with $H_{\text{env}}^{\text{dipole-2qubit}}$ and $H_I^{\text{dipole-2qubit}}$ defined in Eq. \ref{eq:dp2qb}. We focus on the dipole-dipole coupled TLS for two qubit controls. We have set $E_0 \neq E_1$ here since this is required for the ansatz to be universal according to \cite{Morales2020}. The $\sigma^z$ terms resemble the Hamiltonian of a CPHASE gate, practically relevant to superconducting devices. \cite{Krantz2019}\\
    
    We note that for three or more qubits, this ansatz is only universal for an odd number of qubits~\cite{Morales2020}, due to the symmetry of the coupling constants when the number of qubits is even. However, this is not true when there are only two qubits, since there is only one qubit-qubit coupling parameter, and the two-qubit ansatz is actually universal in this special case. A detailed proof is given in Appendix \ref{sec:2qb_uni_proof}. This is also validated by our numerical experiments, where we find that a fidelity of seven nines can be achieved by this control ansatz for 2 qubits alone, i.e., in the absence of coupling to TLS.

	\section{Methods} \label{sec:Methods}
	\subsection{Theoretical Methods}
	\subsubsection{Unitary Fidelity}
	An intuitive fidelity measure is the overlap between the final state, as the result of control, and the target state. However, the states of the bath spins or TLS are not of interest, so a fidelity measure assuming an optimal bath condition has been proposed in \cite{Grace2010,Nielsen2010,Kosut2019}. 
	
	Ref. \cite{Grace2010} first defines a distance between the unitary evolution in Eq. \ref{eq:UniQAOA} and the tensor product between the target gate on the qubits $W$ and an arbitrary unitary $\Phi$ on the bath spins by taking the Frobenius norm of the difference matrix. Assuming that the optimal unitary factorizes, this yields  
	\begin{equation}
	D(\boldsymbol{\theta}, \Phi)=\|U(\boldsymbol{\theta})-W \otimes \Phi\|_{\mathrm{fro}}^{2}.
	\end{equation}
    The optimal control will minimize this distance. However, the choice of the bath unitary $\Phi$ will still affect the distance $D(\boldsymbol{\theta})$. To eliminate this bath dependence, another fidelity measure $F(\boldsymbol{\theta})$ w.r.t optimal $\Phi$ is defined \cite{Grace2010}, which for a single qubit or two qubits interacting with the bath spins is
	 \begin{equation}
	 \min _{\Phi \in \mathrm{U}\left(N_{B}\right)} D(\boldsymbol{\theta}, \Phi) = 2 N(1-\sqrt{F(\boldsymbol{\theta})}).
	 \end{equation}
	Here $N_B = 2^n$ is the total number of degrees of freedom of the bath, and $N = 2^{n+1}$ or $N = 2^{n+2}$ is the total number of degrees of freedom of the system and bath. This optimization can be done analytically \cite{Grace2010} to obtain an optimal bath unitary $\Phi$ that is characterized by fidelity 
	\begin{equation}\label{eq:AUfid}
	F(\boldsymbol{\theta}) = ( \operatorname{tr}\{\sqrt{Q^{\dagger} Q}\} /N ) ^2,
	\end{equation}
	where $Q(\boldsymbol{\theta}) = \operatorname{tr}_{S}\left\{\left(W \otimes I_{\mathrm{bath}}\right)^{\dagger} U(\boldsymbol{\theta})\right\}$. In this work we classically optimize the fidelity measure in Eq. \ref{eq:AUfid} to obtain high-fidelity gate implementations. 
	
	Th fidelity of Eq. \ref{eq:AUfid} is state-independent, with the target being a system unitary. Eq. \ref{eq:uni} presents the target unitary gate matrices $W$ which we use here for benchmarking the bipartite control method with different physical systems. Together these generate the Clifford + T universal set of gates \cite{Gottesman1998}.
	\begin{equation}\label{eq:uni}
	     Z \equiv \sigma^z = \begin{pmatrix} 1 & 0\\ 0 & -1 \end{pmatrix} \quad 
	     \text{Hadamard} \equiv \frac{1}{\sqrt{2}}\begin{pmatrix} 1 & 1\\ 1 &-1 \end{pmatrix} \quad 
	     \text{T} \equiv \begin{pmatrix} 1 & 0\\ 0 & e^{i\pi/4} \end{pmatrix} \quad 
	     \text{CNOT} \equiv \begin{pmatrix} 1 & 0 & 0 & 0\\ 0 & 1 & 0 & 0\\ 0 & 0 & 0 & 1\\ 0 & 0 & 1 & 0\\ \end{pmatrix}
	\end{equation}
	\subsubsection{Average State Fidelity}
	While the fidelity measure from Eq. \ref{eq:AUfid} is state-independent, it is often practically relevant to check how the controlled final state and the target state overlap for a given gate implementation. For this purpose, we checked the optimal protocols obtained from optimization using a state fidelity defined as the average over $M$ initial states sampled randomly and uniformly, i.e.
	\begin{equation}\label{eq:avg_state_fid}
	    F_S (\boldsymbol{\theta}) = \frac{1}{M} \sum_m^M \bra{\psi_T} \rho_{Sm}(\boldsymbol{\theta}) \ket{\psi_T}.
	\end{equation}
	The time-evolved reduced density matrix at time $T$ is defined as
	\begin{equation}
	    \rho_{Sm}(\boldsymbol{\theta}) = \text{tr}_B (U(\boldsymbol{\theta}) \ket{\psi_{im}} \bra{\psi_{im}} U(\boldsymbol{\theta})^\dagger ),
	\end{equation}
	where $\ket{\psi_{im}} = \ket{m_S} \otimes \ket{m_B}$, with $\ket{m_S}$ and $\ket{m_B}$ being Haar-random states of the system and the bath respectively, and $\ket{\psi_T} = W\ket{m_S}$ is the target state. The Haar-random states are generated using the Python software qutip \cite{JOHANSSON20131234}.
	
	\subsubsection{Fidelity with respect to a set of reference states}
	In the case of Lindblad evolution, the unitary fidelity of Eq. \ref{eq:AUfid} is not applicable, as the Lindblad evolution provides a quantum channel instead of a unitary over the qubits and coupled spins that constitute the primary bath. Instead, we use the fidelity measure proposed in \cite{Goerz_2014} for optimizations with Lindblad dynamics. This fidelity is the average of the Hilbert–Schmidt product between the target state and the evolved density matrix with respect to a set of reference states \cite{Goerz_2014, Koch2016},
	\begin{equation} \label{eq:ref_state_fid}
	    	F(\boldsymbol{\theta}) = \sum_{i=0}^d \frac{w_i}{\text{Tr}\left[\rho_i^2\right]} \Re e \left\{ \text{Tr}\left[ W\rho_iW^{\dagger} \mathcal{D}[\rho_i] \right] \right\}.
	\end{equation}
    Here $d+1$ is the number of reference states in the set, $\rho_i$ is the $i$-th reference state, $w_i$ is a weight for each reference state with $\sum_i w_i = 1$ and 
    \begin{equation}
        \mathcal{D}[\rho_i] = \text{tr}_B \left(\mathcal{L}[\rho_i]\right)
    \end{equation}
    is the quantum channel governing the evolution. $\mathcal{L}[\cdot]$ is the Lindblad evolution of Eq. \ref{eq:Lind} for the qubits and primary TLS bath.
    
  Following \cite{Goerz_2014}, for efficiency and accuracy we choose $d= 2$ for a single qubit and $d=4$ for two qubits, corresponding to the dimension of the system Hilbert space in each case. The reference states are $\rho_i = \ket{\varphi_i}\bra{\varphi_i}$ for $i = 0,\dots, d-1$, with $\ket{\varphi_i}\bra{\varphi_i}$ being computational basis states and $(\rho_d)_{ij} = \frac{1}{d}$. All weights are set equal to $w_i = \frac{1}{d+1}$. The TLS initial state is the zero temperature state. 

	\subsection{Numerical Methods}
	\subsubsection{Optimization Algorithms} \label{sec:OptAlg}
    Optimization of the bipartite control problem is a numerically challenging task and requires careful design of the optimization algorithm \cite{Arenz2014,Kosut2019}. In this work, we employ reinforcement learning with \textit{policy gradient} (PG), which have been shown to be effective for optimizing QAOA protocols on systems with classical noise. \cite{Yao2020,Yao2020Robust,Yao2020cd}. We also applied the sequential convex programming (SCP) method of \cite{KosutGB:13,Dong2019} which showed outcomes similar to those obtained with PG. To keep the focus on the Hamiltonian switching protocol rather than on the algorithmic implementations, we present only the PG results here.
    
    The PG method used in this work was developed in \cite{Yao2020,Yao2020Robust,Yao2020cd}. The key idea is to treat the values each parameter as a probability distribution and to then optimize over the mean and standard deviation of the distribution, instead of over the parameter themselves. This has the advantage of not requiring any information about the derivatives of the fidelity function. The final result is expected to converge to a delta distribution with the mean value being the optimal value of the parameter \cite{Yao2020}.
    The original PG method developed in \cite{Yao2020} has the disadvantages of unbounded protocol durations and possibility of negative time intervals, which limits its application to Hamiltonian switching control. These issues were fixed in subsequent work \cite{Yao2020Robust,Yao2020cd} and we use here the improved versions of the PG methods from Refs. \cite{Yao2020Robust,Yao2020cd}.
	
	Although we can set constraints on the time duration of the protocols and avoid the occurrence of negative entries (i.e., non-physical negative time intervals), small negative values are nevertheless still observed in optimizations. This issue is resolved by taking the absolute value of the negative entries in the protocol when evaluating the fidelity. The number of PG iterations is set to 2000 to ensure convergence (see Sec. \ref{sec:trials}). All simulations are carried out with several optimizations running in parallel. To achieve the best result within a reasonable simulation time, each simulation is run 3 to 5 times with different initial interval sets. In all cases we show the best results from all of the parallel simulations.

    When the accuracy reached with PG optimization of Hamiltonian switching is not sufficient, we conduct a secondary optimization using the GRAPE method \cite{Khaneja2005} on the outcomes of Hamiltonian switching control. Instead of optimizing the hold times as in the Hamiltonian switching control, we now optimize the amplitude of each control pulse in GRAPE with their time durations fixed at the values obtained from the Hamiltonian switching optimization. The numerical optimization in GRAPE is carried out using gradient-based methods.
	
	\subsubsection{Numerical Details}
    The unitary dynamics are implemented using NumPy \cite{harris2020array}. The unitary evolution operator Eq. \ref{eq:UniQAOA} is generated by directly multiplying $-iH_A$ and $-iH_B$ in computational basis with time durations $\alpha_i,\beta_i$, exponentiating the resultant matrix, then multiplying all $2p$ matrices together. Simulation of Lindblad dynamics is carried out with the superoperator approach implemented using NumPy \cite{harris2020array} for three or less total number of spins and TLS (system plus primary bath), and by numerical integration of the ordinary differential equation implemented with qutip \cite{JOHANSSON20131234} for larger models. All of the Pauli operators in the computational basis are generated using Python package quspin \cite{Weinberg2017}. 
    
    The GRAPE optimizations are done using qutip \cite{JOHANSSON20131234} with modifications. The gradient of fidelity measure Eq. \ref{eq:AUfid} used in GRAPE optimization is given in \cite{Floether_2012}. The optimization method is the BFGS algorithm implemented by scipy \cite{2020SciPy-NMeth}. The control amplitudes are constrained to lie within the range $[-1.2, 1.2]$, while the control amplitudes in Hamiltonian switching control are held at $\pm 1$.
    
    In all of the plots shown in this work, the fidelity is presented as \textit{minus log infidelity} (MLI), i.e., the vertical axis of the displayed plots
    \begin{equation}\label{eq:MLI}
        -\log_{10}(1-F(\theta^*)).
    \end{equation}
    This allows the fidelity differences between different simulations to be better assessed, since most of the fidelities are close to 1. The integer part of the MLI gives the number of nines in the actual fidelity.
    
	In numerical experiments for a certain physical model, we focus on characterizing the change of the fidelity with the control depth $p$ and the total evolution time $T$. The control depth $p$ corresponds to the number of parameters for the circuit parameterization: larger $p$ gives added flexibility to the protocol, which is expected to give more powerful protocols and thus higher fidelity. However, larger control depth adds to the effort of optimization and is experimentally harder to implement. It is also desirable for the total evolution time $T$ to be short for fast gate implementation. However, for an ansatz with fixed control strength there is a theoretical lower bound for the required gate time ~\cite{Bukov2018,Hegerfeldt2013}. Since any protocol with shorter gate time would give lower fidelity, we therefore simultaneously probe the effect of both $p$ and $T$ on the fidelity of the Hamiltonian switching control, and look for low-depth and short-time protocols that give fidelity values larger than 0.9999, corresponding to infidelities, i.e., gate errors, below the threshold of most contemporary quantum error correction codes. \cite{Terhal2015,Aliferis2007,Campbell2017}\\
	\subsubsection{Trial Simulations}\label{sec:trials}
\begin{figure}[h!]
	 	\centering
	 	\begin{subfigure}[b]{\figwidthThird}
	 	     \includegraphics[width=\linewidth]{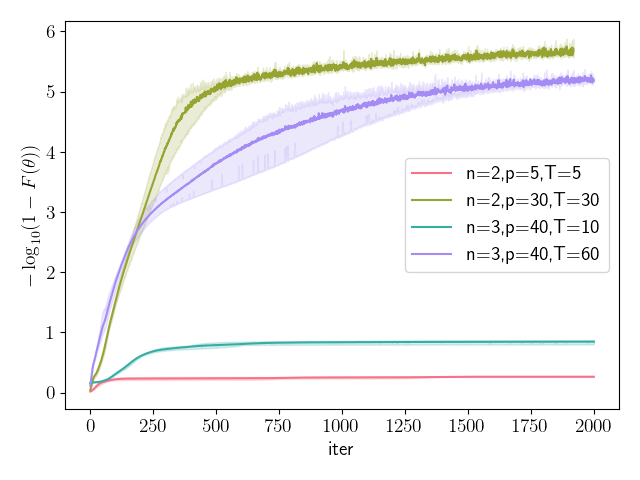}
	 	     \caption{}
	 	     \label{fig:fid_iter}
	 	\end{subfigure}
	 	\begin{subfigure}[b]{\figwidthThird}
	 	     \includegraphics[width=\linewidth]{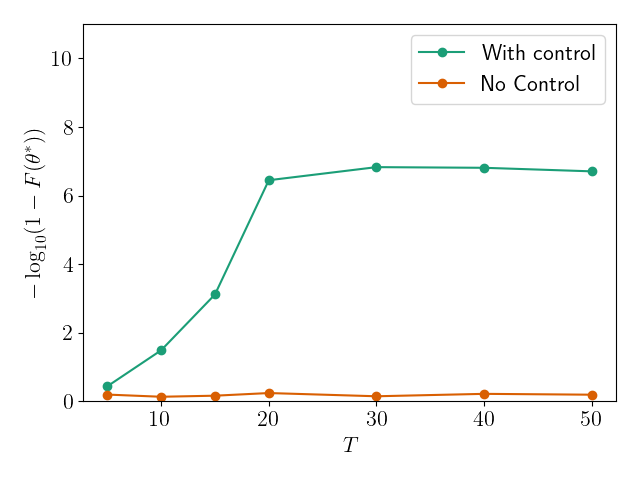}
	 	     \caption{}
       		\label{fig:iter_comparison}
	 	\end{subfigure}
	 	\caption{(a) Dependence of MLI of fidelity (Eqs. \ref{eq:AUfid} and \ref{eq:MLI}) on the number of iterations with PG, for a $Z$ gate on a single qubit that is isotropically coupled (Eq. \ref{eq:CSnoCtr}) to $n$ environmental spins.  Run parameters are displayed in the legend. 
        (b) Comparison of the unitary fidelity (Eq. \ref{eq:AUfid}) for a Z gate on a single qubit isotropically coupled to $n=2$ bath spins under evolution with (green curve)or without (orange curve) Hamiltonian switching.
        The horizontal $T$ axis is the total duration of evolution for the bipartite protocol. The control depth for the green curve is $p=20$ and the optimization method used is PG. The fidelities of the orange points without Hamiltonian switching  are calculated with the same total duration of the corresponding control protocol and would be 1 if there were no decoherence due to coupling to the primary bath of spins.
        }
	\end{figure}

	To find the number of iterations required to achieve optimal fidelity with reasonable resources in a PG optimization, a series of test optimizations were first carried out. Fig. \ref{fig:fid_iter} shows the MLI dependence on the number of PG iterations for a $Z$ gate on a single qubit isotropically coupled (see Eq. \ref{eq:CSnoCtr}) to $n$ environmental spins. We find that 2000 iterations are enough for the optimizer to converge. The vertical axis here is the best fidelity reached in any single run. The uncertainties are due to averaging over several (3 or 5) simulations with different initial protocols. It is evident that the fidelity is not significantly improving after 1000 iterations, so 2000 is enough to exploit the power of the optimizer. The fidelity change over the number of iterations for other simulations are also checked and show a similar trend, confirming that 2000 iterations is a valid choice.\\
	
	We also compare with the fidelity for evolution without any control. Fig. \ref{fig:iter_comparison}) shows the MLI as a function of the control protocol time $T$ for a Z gate on a single qubit isotropically coupled to $n=2$ bath spins under evolution with (green curve)or without (orange curve) Hamiltonian switching. 
    This comparison is made with a state evolved with the same number of bath spins under the Hamiltonian that directly produces the unitary within the total duration of the protocol $T$. For example, for the target gate $Z$, the system part of the Hamiltonian is $H_s' = -\frac{\pi}{2T}\sigma^z$, which gives $\exp(-iH_s'T) = \sigma^z$ up to phase. Then the unitary under decoherence due to the primary bath is given by $U_n = \exp(-i(H_s' +H_I)T)$, and the fidelity under this evolution is given by the fidelity measure in Eq. \ref{eq:AUfid}. In the absence of coupling to the bath ($H_I=0$), both of these evolutions would give unit fidelity. However, it is evident from Fig. \ref{fig:iter_comparison}) that the fidelity is significantly lower in the presence of coupling to the bath. Implementing the Hamiltonian switching bipartite control leads to a considerable improvement in fidelity,  
    showing the effectiveness of the Hamiltonian switching bipartite control.\\
	
\section{Results} \label{sec:results}

    \subsection{Single qubit} \label{sec:CS}
    \subsubsection{Isotropic coupling} \label{sec:single_iso}
\begin{figure}[h!]
	 	\centering
	 	\begin{subfigure}[b]{\figwidthThird}
	 	     \includegraphics[width=\linewidth]{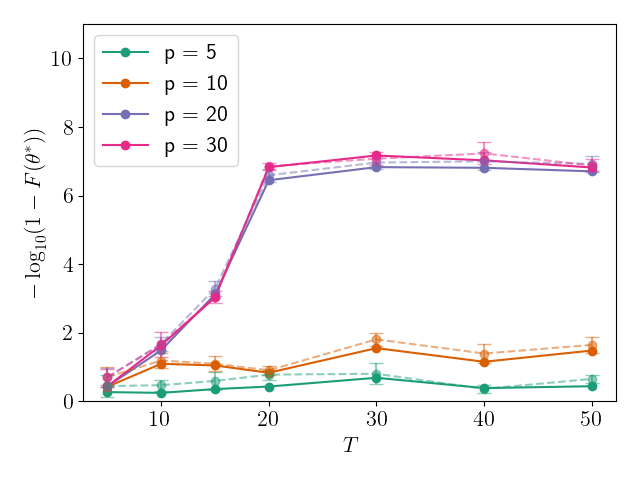}
	 	     \caption{$n=2$, fixed $p$}
	 	     \label{fig:CSpgn2}
	 	\end{subfigure}
	 	\begin{subfigure}[b]{\figwidthThird}
	 	     \includegraphics[width=\linewidth]{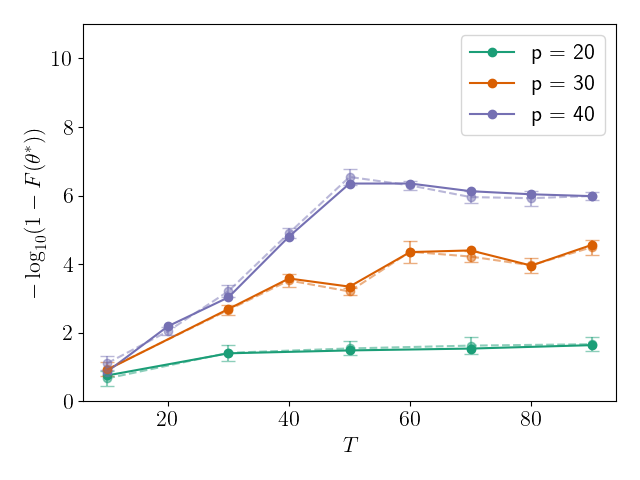}
	 	     \caption{$n=3$, fixed $p$}
	 	\end{subfigure}
	 	\begin{subfigure}[b]{\figwidthThird}
	 	    \includegraphics[width=\linewidth]{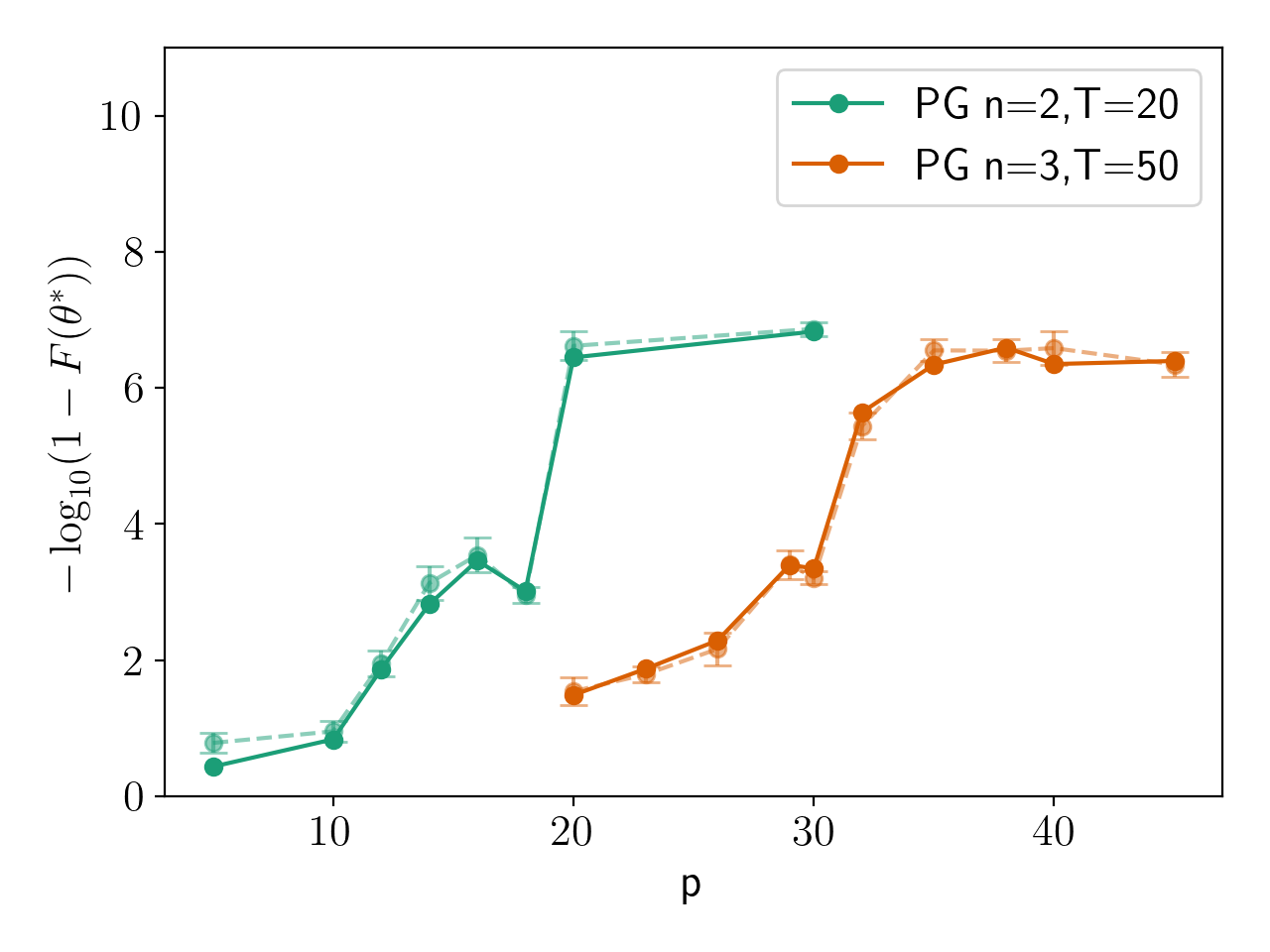}
    	 	\caption{$n=2,3$, fixed $T$}
    	 	\label{fig:cs_p}
	 	\end{subfigure}
	 	\caption{$Z$ gate fidelity dependence on total evolution time T ( panels (a),(b)) and control depth p (panel (c)) for isotropically coupled systems optimized with PG. All coupling constants are equal to $A_q=1.0$. Each point is the best result of 3 or 5 parallel simulations with different protocol parameter initialization. For each pair of points with the same color and T value, the solid line is the actual optimization result and the dashed pale line with error bar is the average state fidelity over $M=100$ Haar-random states. The fidelity, Eq. \ref{eq:AUfid}, is plotted as MLI, Eq. \ref{eq:MLI}.}
 		\label{fig:szn23}
	\end{figure}

    We first study control of the isotropically coupled central spin system with all the coupling constants set equal to $A=1$. We start with the target gate $Z$. The optimization results for $n=2,3$ bath spins are displayed in Fig. \ref{fig:szn23}.
	
	In all of the test-cases, the highest fidelity reached is above $1-10^{-6}$, given sufficient control depth $p$ and evolution time $T$. A notable feature is that when the protocol depth $p$ is below a certain threshold value, the fidelity does not change significantly with $T$. For instance, in Fig. \ref{fig:CSpgn2}, the fidelity has no significant change with $T$ when $p<20$, but for $p\geq20$, it increases drastically when $T$ changes from 10 to 20, displaying the phase-transition behavior observed in \cite{Bukov2018}. We refer to the minimal depth required for significant fidelity improvement with time as the \textit{critical depth} $p^*$. Beyond the critical depth $p^*$, the fidelity generally first increases with total evolution time $T$ and then reaches a plateau. We refer to the transition time between the increasing and plateau regions for given number of spins $n$ as the \textit{critical time} $T^*$. Panel (a) for $n=2$ shows that in the plateau region, the fidelity may not improve much, as the depth $p$ further increases. To better reveal the fidelity dependence on protocol depth $p$, Fig. \ref{fig:cs_p} shows the fidelity plotted as a function of $p$, for fixed time $T$. For $n=2$ we see a significant improvement in fidelity as $p$ increases to 20, and we can infer that $p^* = 20$. For $n=3$ the corresponding jump is not so drastic, but we can still see the appearance of a critical depth at $p^*=32$. In general, the fidelity changes with time and with control depth display similar behaviors, with both increasing to a plateau value. The critical points represent the onset of plateau for each parameter. We note that only when one parameter is located beyond its critical value, can the other parameter show this pattern of increase to a plateau value. In all other situations the fidelity is restricted to low values.
    
    In these calculations the optimization was carried out with the unitary fidelity of Eq. \ref{eq:AUfid}. The corresponding state fidelities of Eq. \ref{eq:avg_state_fid} are also plotted in Fig. \ref{fig:szn23}, as dashed pale lines of the same color. The size of the Haar-random sampling of states is set to $M=100$: this value produces an almost identical mean and standard deviation as the larger values $M=1000$ and $M=10000$. In all the cases, the standard deviation of the state fidelities are very low, showing only a weak dependence on the initial states. The MLI differences between mean state fidelity and unitary fidelity are all within 0.5, corresponding to fidelity differences less than $10^{-7}$ in the plateau region, and the two fidelities display almost identical behavior. In general, we thus find that the unitary fidelity is an effective generalized fidelity measure for optimization.
    
    With knowledge of the critical depth $p^*$ for the target gate $Z$, we can now choose parameters that give high fidelity with a relatively low depth and use these to test the critical depth for Hadamard and T gates. We chose $n=2, p=20,30$, which are just above the critical depth $p^*=20$.

	\begin{figure}[h]
	 	\centering
	 	\begin{subfigure}[b]{\figwidthThird}
	 	    \includegraphics[width=\linewidth]{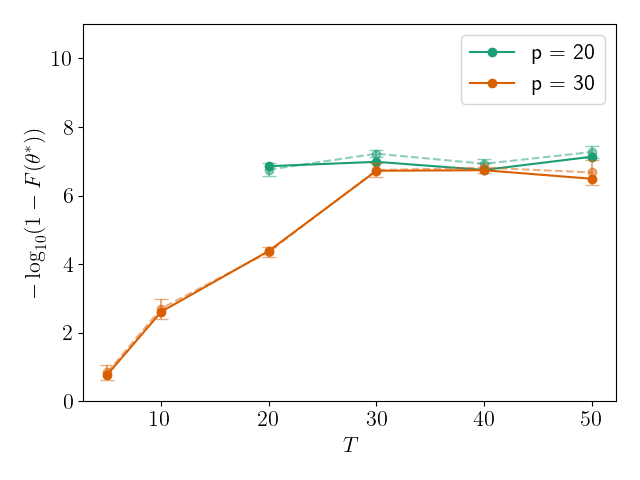}
    	 	\caption{Hadamard Gate}
    	 	\label{fig:Heis_Had}
	 	\end{subfigure}
	 	\begin{subfigure}[b]{\figwidthThird}
	 	     \includegraphics[width=\linewidth]{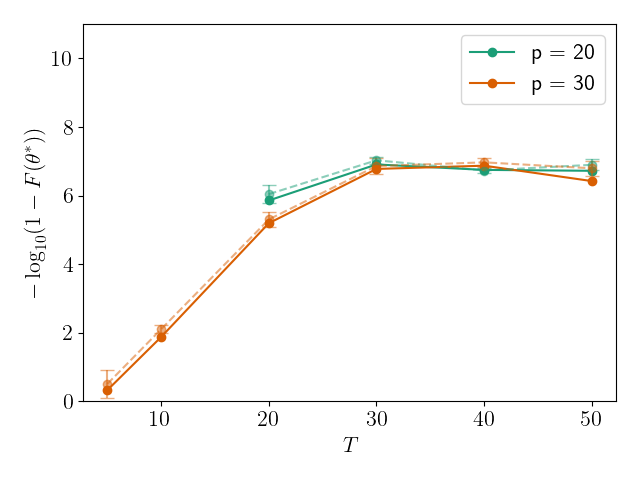}
	 	     \caption{T gate}
	 	     \label{fig:Heis_T}
	 	\end{subfigure}
	 	\begin{subfigure}[b]{\figwidthThird}
	 	     \includegraphics[width=\linewidth]{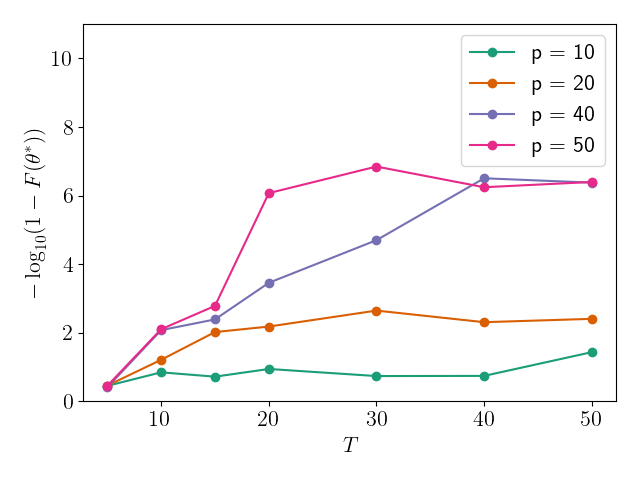}
	 	     \caption{Different Coupling Strengths, $Z$ gate}
	 	     \label{fig:Heis_uneq}
	 	\end{subfigure}
	 	\caption{Single qubit gate fidelity dependence on total evolution time for isotropically coupled central spin systems with $n=2$ bath spins optimized with PG. The target gates are (a) Hadamard gate, (b) T gate and (c) $Z$ gate. In panels (a), (b), the coupling constants are all equal to $A_q=1.0$. In panel (c), the coupling strengths are unequal and evenly distributed in the range $[1,2]$. Each point in (a) and (b) is the best result of 3 or 5 parallel simulations with different initialization. For each pair of points with the same color and T value, the solid line is the actual optimization result and the pale dashed line with error bar is the average state fidelity over $M=100$ Haar-random states. All fidelities are plotted as MLI, Eq. \ref{eq:MLI}.}
	 	\label{fig:Heis_HadTDiff}
	\end{figure}
	The fidelity dependence on time $T$ for the Hadamard gate and T gate targets are plotted in Fig. \ref{fig:Heis_Had} and \ref{fig:Heis_T} . These show similar trends to the results with the $Z$ gate in Fig. \ref{fig:szn23}. The fidelities reached for each of these target gates are of similar magnitude at sufficiently large $T$ and $p$, with more than six nines in most  cases. The critical depths and critical times are about $p^* = 20$ and $T^* = 20$, which are the same as the $Z$ gate. Thus, we can conclude that the control properties for different 1-qubit target unitaries are similar.
	
	The results for the isotropic coupling model with equal coupling constants, i.e., all $A_q=1$, are summarized in Table \ref{tab:Heis}. Note that the results for $n=1,4$ are not plotted but the patterns are similar to those found in Figs. \ref{fig:szn23} and \ref{fig:Heis_HadTDiff}.\\
    \begin{table}[h]
	    \centering
	    \begin{tabular}{c|c|c|c|c}
	       Target gate & \# of bath spins ($n$) & Critical depth ($p^*$) & Critical time ($T^*$) & Highest MLI\\
	       \hline
	       \multirow{4}{*}{$Z$} & 1 & $\sim$10 & $\sim$10 & 8.14\\ \cline{2-5}
	                           & 2 & $\sim$20 & $\sim$20 & 7.17\\ \cline{2-5}
	                           & 3 & $\sim$32 & $\sim$40& 6.35\\ \cline{2-5} 
	                           & 4 & $\sim$60 & $\sim$70& 5.38\\ \hline
        	Hadamard& 2 & $\leq 20$ & $\sim$20 & 6.63\\ \hline
        	T & 2 & $\leq 20$ & $\sim$20 & 6.91\\ \hline
	    \end{tabular}
	    \caption{The estimated critical depth $p^*$ and critical time $T^*$ values, together with the highest log fidelity reached in the simulations of the central spin model with isotropic couplings and equal coupling strengths.}
	    \label{tab:Heis}
	\end{table}
	
	While not exact, the critical value assignments in Table \ref{tab:Heis} already show rich features of the control. A general trend with increasing bath spin count $n$ is that the critical time $T^*$ and critical depth $p^*$ increase, while the highest fidelity drops. This trend is reasonable, since an increased bath size increases the noise strength and adds to the difficulty of the control task. This result is consistent with the reduced dynamics shown in Fig. \ref{fig:Heis_dyna} in the absence of the control Hamiltonian, in which the dynamics of systems with different bath spin count behave very differently within the timescale of the protocols.

    As shown in \cite{Arenz2014}, the controllability is rather different when the coupling constants between the system and bath spins $A_q$ vary with $q$. Specifically, when the coupling constants are different, the bath spins of the central spin model with isotropic couplings become controllable through bipartite control. However, the additional controllable degree of freedom adds difficulty to the control problem, and was found in \cite{Arenz2014} to result in lower fidelity than the equal coupling case. Fig. \ref{fig:Heis_uneq} shows the result of applying Hamiltonian switching on a system with the same range of coupling strengths as that in \cite{Arenz2014}, which is uniformly distributed between 1 and 2.
    We can see from the plots that the difficulty of control is significantly greater when the couplings are different. The critical depth for 2 qubits increases to 40 or 50. In the case of $n=2, p=50$, we see the typical behavior of an initial increase in the MLI, and hence the fidelity, which is followed by a plateau above a critical depth $p^*$. Note that the critical time is still around 20, which is the same as the case of $n=2$ for equal couplings in Fig. \ref{fig:CSpgn2}. Changing to the variable coupling strengths $A_q$ increases the critical depth but appears to keep the critical time $T^*$ the same. Notably, the fidelity above critical time and depth can still reach six nines. 

	\subsubsection{Dipole-dipole Coupling - TLS bath} \label{sec:Transmon}
\begin{figure}[h!]
	 	\centering
	 	\begin{subfigure}[b]{\figwidthThird}
                \includegraphics[width=\linewidth]{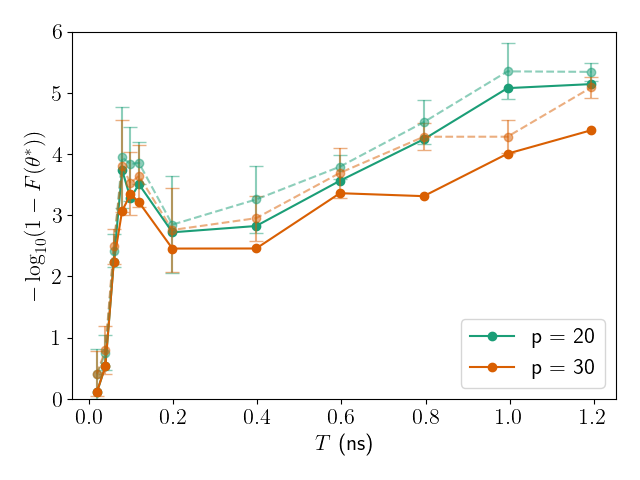}
                \caption{$n=2$, fixed $p$, $Z$ gate}
                \label{fig:TLS_single_n2T}
	 	\end{subfigure}
	 	\begin{subfigure}[b]{\figwidthThird}
	 	    \includegraphics[width=\linewidth]{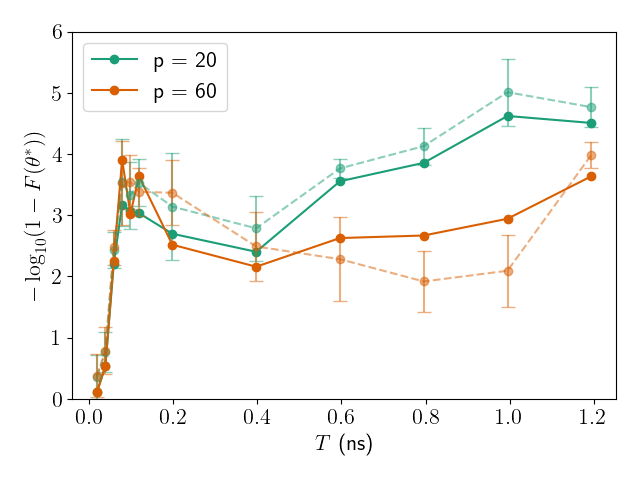}
	 	    \caption{$n=3$, fixed $p$, $Z$ gate}
                \label{fig:TLS_single_n3T}
	 	\end{subfigure}\\
	 	\begin{subfigure}[b]{\figwidthThird}
	 	    \includegraphics[width=\linewidth]{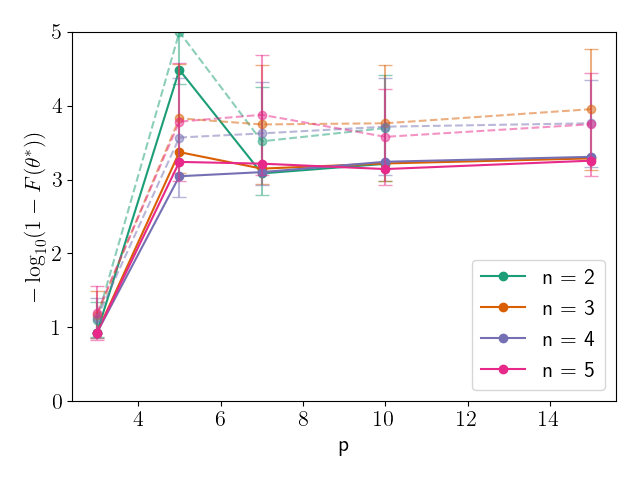}
	 	    \caption{$n=2-5$, fixed $T=\frac{4}{16\pi}$ ns, $Z$ gate}
	 	\end{subfigure}
	 	\begin{subfigure}[b]{\figwidthThird}
	 	    \includegraphics[width=\linewidth]{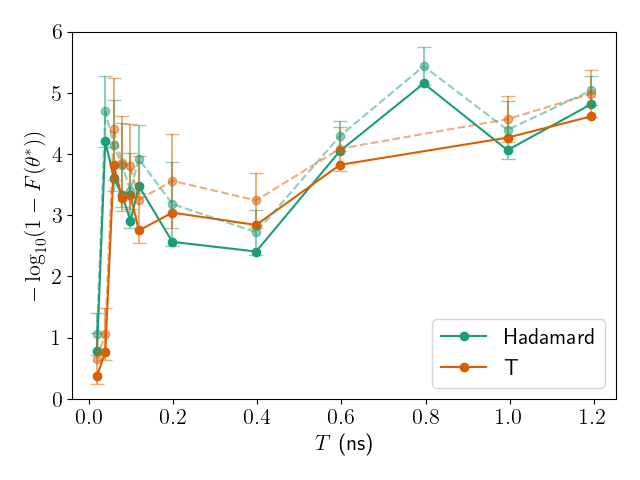}
	 	    \caption{$n=2, p=20$, Hadamard and T gates}
	 	    \label{fig:TLS_single_HadT}
	 	\end{subfigure}
	 	\caption{Single qubit gate fidelity dependence on (a),(b) and (d): total evolution time and (c) control depth for dipole-dipole coupled systems optimized with PG. The target gate is $Z$ for panels (a)-(c) and Hadamard and T for panel (d). The coupling constants are unequal and evenly distributed over 4-40 MHz, which is in the relevant strength range for superconducting devices. Each point is the best result of 3 or 5 parallel simulations with different initialization. For each pair of lines with the same color and T value, the solid one is the actual optimization result and the pale dashed one with error bar is the average state fidelity over $M=100$ Haar-random states. All fidelities are plotted as MLI, Eq. \ref{eq:MLI}.}
        \label{fig:TLS_single}
	\end{figure}

	Here we apply bipartite Hamiltonian switching control to a single qubit coupled to TLS through dipole-dipole couplings as described in Eq. \ref{eq:TLSint}. In this model we focus on the variable coupling case ($A_q$ sampled uniformly on $[4,40]$ MHz) for practical relevance, since in superconducting qubits the coupling strengths between the qubit and individual TLS are generally variable \cite{Simmonds2004,Neeley2008,Cole2010}. The results of fidelity dependence on time with PG optimization are shown in Fig. \ref{fig:TLS_single_n2T}, \ref{fig:TLS_single_n3T} for $n=2,3$ bath spins. Notice that the horizontal axis is now in units of nanoseconds. One time unit in these simulations, corresponding to  the time unit in the isotropic model of Sec.~\ref{sec:single_iso} above, is now $\frac{1}{16\pi} \approx 0.02$ ns. The fidelity dependence on $T$ for $n=1,4,5$ is also simulated and shows very similar behavior. Trial simulations with all coupling constants equal to 40MHz yield results very similar those shown in Fig. \ref{fig:TLS_single}.\\

	The fidelity dependence on $T$ and $p$ for all simulations are rather similar. The critical time for all the $Z$ gate simulations is $T^* = \frac{4}{16\pi}\, \text{ns} \approx 0.08$ ns (4 time units), which is significantly shorter than the corresponding value for the isotropically coupled model in the previous section. The critical depth is also much lower than for the central spin model with isotropic couplings, with all simulations for $n=2-5$ displaying the pattern of an increase to a plateau value for $p=20$ that does not change significantly for larger $p$. Additional simulations suggest that the critical depth for all $n=2-5$ test cases with target gate $Z$ have critical depth $p^*=5$. The significant decrease of critical time and critical depth seen relative to the values for the isotropically coupled model is likely the result of a much weaker coupling strength compared to that used for the calculations with the central spin model and isotropic couplings in Sec. \ref{sec:CS}. This is confirmed by trial simulations carried out with dipole-dipole coupled qubits having the same coupling strengths as those employed in the isotropic model (i.e., of the same order of magnitude as the system energy splitting). These simulations now give  similar $T^*$ and $p^*$ values to those found for the isotropically coupled model. 
	
	The results for Hadamard and T gates are shown in Fig. \ref{fig:TLS_single_HadT}. The critical time of the Hadamard gate is $T^* = \frac{2}{16\pi}\, \text{ns} \approx 0.04$ ns (2 time units) and of the T gate is $T^* = \frac{3}{16\pi}\, \text{ns} \approx 0.06$ ns (3 time units), which are both shorter than the $Z$ gate. The critical depths are also lower. The highest fidelity achieved is also around four nines. In general, the control properties of different target gates are quite similar, but the Hadamard and T gates are expected to be easier to implement because the critical depths are lower and the critical times are shorter compared to the $Z$ gate.
	
	In general, all of the curves for the TLS coupling model show a rather similar pattern, with a short critical time $T^*$ and a low critical depth $p^*$, and with final gate fidelity limited at around four nines. The results are summarized in Table. \ref{tab:TLS}.
    \begin{table}[h]
	    \centering
	    \begin{tabular}{c|c|c|c}
	       Target gate  & \# of bath spins ($n$) & Critical depth ($p^*$, $T=T^*$) & Highest MLI\\
	       \hline
	       \multirow{4}{*}{$Z$} & 2 & 5 & 5.14 (9.48)\footnote{Highest MLI obtained with secondary GRAPE optimization.} \\ \cline{2-4}
	                           & 3 & 5 & 4.62 \\ \cline{2-4}
	                           & 4 & 5 & 3.66\\ \cline{2-4} 
	                           & 5 & 5 & 3.66\\ \hline
	        Hadamard & 2 & $\leq 3$ & 5.15\\ \hline
	        T & 2 & $\leq 3$ & 4.61\\ \hline
	    \end{tabular}
	    \caption{The estimated critical depth $p^*$ and highest log fidelity reached in the simulations of dipole-dipole couplings.}
	    \label{tab:TLS}
	\end{table}
	The similarity of critical time and depth among different numbers of TLS agree with the characteristic feature of the reduced dynamics of the control-free system shown in Fig. \ref{fig:dp_dyna}, namely, that the dynamics of a single qubit coupled to different number of TLS appear almost identical within the timescale of the simulations.
	
    We note that the state fidelity of the protocols for the TLS model is somewhat higher than the gate fidelity. The standard deviation of the state fidelity is also larger than that for the isotropic coupling model, but are still generally small, indicating weak fidelity dependence on initial states.

    It is apparent that for qubits with dipolar coupling to the primary bath spins, Hamiltonian switching alone does not always reliably increase the fidelity significantly above the desired fidelity. To further improve the fidelity, we applied GRAPE to undertake a secondary optimization of the piecewise constant control amplitudes with the duration of each piece fixed to the optimal protocol derived from the optimal Hamiltonian switching control. This was carried out on the results for $n=2$ bath spins with $T=20,30,40,50,60$ time units (approximately 0.4, 0.6, 0.8, 1.0 and 1.2 ns) and $p=20$ that are presented in Fig. \ref{fig:TLS_single_n2T}. The MLI after PG optimization are 2.83, 3.56, 4.24, 5.08 and 5.14. After GRAPE optimization, the MLI are improved to 7.65, 8.16, 8.44, 9.48 and 8.45 respectively. These values are now well above our target of four nines. 
 
    \subsubsection{Dipolar CS bath and Lindblad bath}
\begin{figure}[h!]
	 	\centering
	 	\begin{subfigure}[b]{\figwidthHalf}
	 	    \includegraphics[width=\linewidth]{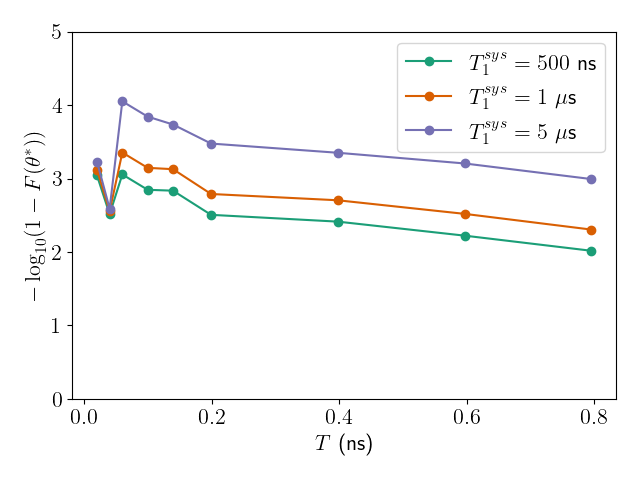}
	 	    \caption{$n=0$ with $T_1$ decay}
	 	    \label{fig:TLS_Lind_n0}
	 	\end{subfigure}
	 	\begin{subfigure}[b]{\figwidthHalf}
	 	    \includegraphics[width=\linewidth]{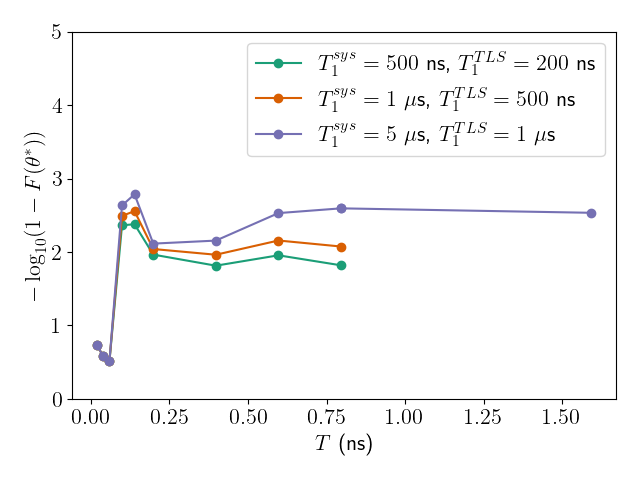}
	 	    \caption{$n=3$ with $T_1$ decay}
	 	    \label{fig:TLS_Lind_n3}
	 	\end{subfigure}
	 	\caption{T gate fidelity dependence on total evolution time for a single qubit coupled to $n$ TLS and optimized with PG. In addition to coupling to the TLS, both the system and TLS are under $T_1$ decay. The coupling constants are unequal and evenly distributed over 4-40 MHz, which is in the relevant strength range for superconducting devices. Each point is the best result of 3 or 5 parallel simulations with different protocol parameter initialization. The fidelity, Eq. \ref{eq:ref_state_fid}, is plotted as MLI, Eq. \ref{eq:MLI}. }
 		\label{fig:TLS_Lind}
	\end{figure}

   We also studied the effect of $T_1$ decay on both the qubit and TLS, as described in Eq. \ref{eq:Lind}. The decay rates of the qubit and all TLS are set to be equal. Results are shown in Fig. \ref{fig:TLS_Lind} for a target T gate on a single qubit, with zero TLS (panel (a)) and with $n=3$ TLS (panel (b)). Simulation costs restricted the PG iterations to 250 in the latter case, but the fidelities  were well converged at this point. Calculations for other gates show similar performance. The fidelity measure used here is given by Eq. \ref{eq:ref_state_fid}. Note that this fidelity measure is defined on a set of reference states that effectively represents an average over all states in the qubit Hilbert space \cite{Goerz_2014}, so we do not additionally present an average state fidelity over Haar-random states.
    
    Fig. \ref{fig:TLS_Lind_n0} shows that spontaneous emission severely reduces the gate fidelity of an isolated qubit ($n=0$), lowering this from a value above ten nines in the absence of both TLS and spontaneous emission, to a value of only three to four nines. The critical behavior with respect to time is still apparent but there is now no real plateau and instead the fidelity (MLI) shows a slow decay over duration of the protocol which is consistent with the dissipative nature of spontaneous emission. Fig. \ref{fig:TLS_Lind_n3} shows that after adding the additional coupling to the TLS, the fidelity further drops, as expected because of the increased overall noise strength with coupling to both the TLS and Lindblad bath, but the critical time is not changed.

    \subsubsection{Non-universal Control}
	All results in the preceding sections are derived with the universal ansatz $ H_S(t) = E \cdot \left(-\frac{1}{2}\sigma^z \pm 2\sigma^x\right)$ (Eq. \ref{eq:TLS_AB}). As mentioned in Sec. \ref{sec:ansatz}, we have also tested the non-universal ansatz $ H_{Sz}(t) = E \cdot \left(-\frac{1}{2}\sigma^z \pm 2\sigma^z\right)$ (Eq. \ref{eq:non_uni_ansatz}) with Hamiltonian switching.
	
\begin{figure}[h!]
	 	\centering
	 	\begin{subfigure}[b]{0.4\textwidth}
	 	    \includegraphics[width=\linewidth]{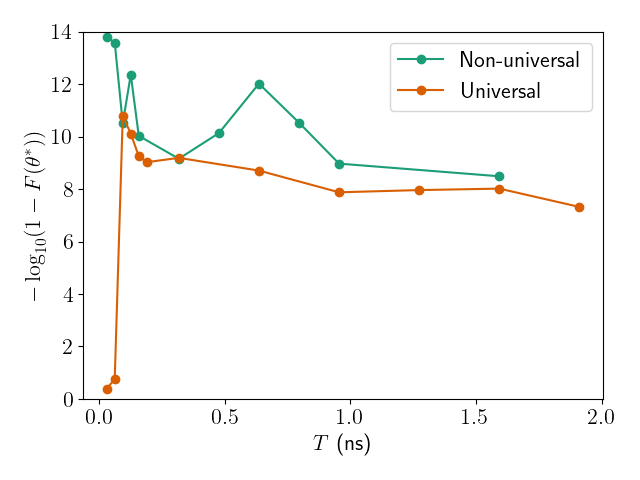}
	 	    \caption{One isolated qubit}
	 	    \label{fig:T_NoBath}
	 	\end{subfigure}
	 	\begin{subfigure}[b]{0.4\textwidth}
	 	    \includegraphics[width=\linewidth]{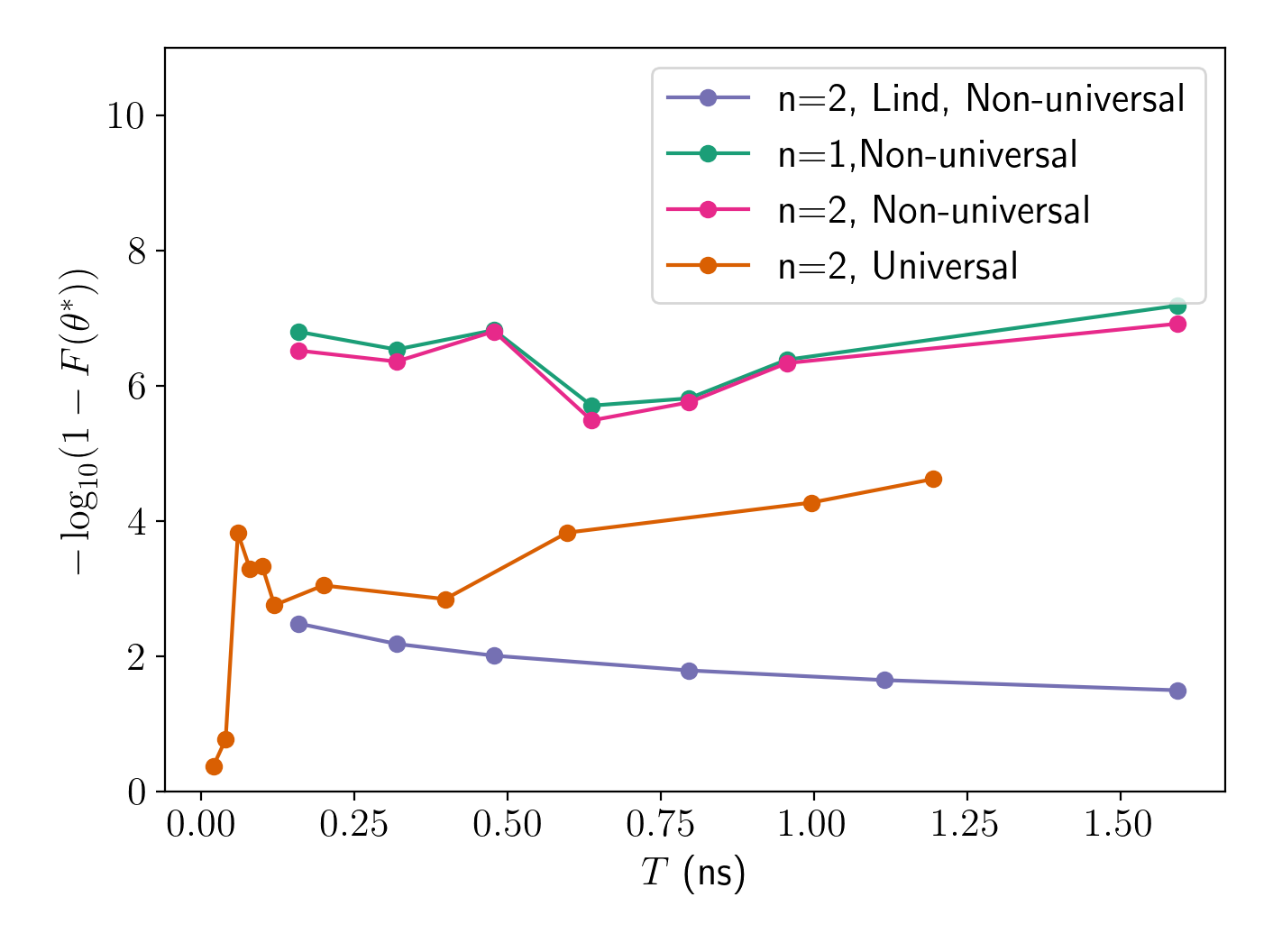}
	 	    \caption{One qubit with TLS and spontaneous emission}
	 	    \label{fig:Non_uni_T}
	 	\end{subfigure}
	 	\caption{T gate fidelity dependence on total evolution time for (a) 1 isolated qubit  (b) 1 qubit dipole coupled to $n$ TLS. Optimizations were performed with PG, using a control depth $p=20$ for calculations without Lindblad decoherence and $p=30$ for calculations with Lindblad decoherence. The control ansatze are Eq. \ref{eq:TLS_AB} (universal) and Eq. \ref{eq:non_uni_ansatz} (non-universal). In panel (b), the purple line shows results for a system where both the qubit and TLS are subject to $T_1$ decay, with $T_1^{sys} = 500$ ns and $T_1^{TLS} = 200$ ns. All other lines show results with no Lindblad decoherence. The dipolar coupling constants are unequal and evenly distributed over $4 - 40 \, \text{MHz}$.} Each point is the best result of 3 parallel simulations with different initialization. The fidelity, Eq. \ref{eq:ref_state_fid}, is plotted as MLI, Eq. \ref{eq:MLI}.
 		\label{fig:NonUniversal}
	\end{figure}

	We first analyzed the performance of this ansatz on implementation of a T gate. Since the gate itself is a $Z$-rotation, the ansatz is able to generate the gate with high fidelity, as shown in Fig. \ref{fig:NonUniversal}. In Fig. \ref{fig:T_NoBath}, the results are shown for a T gate implemented by both the universal ansatz (Eq. \ref{eq:TLS_AB}) and the non-universal ansatz (Eq. \ref{eq:non_uni_ansatz}) on an isolated qubit. The universal ansatz shows the increase to a plateau value seen in the previous sections, whereas the non-universal ansatz gives very a high fidelity (over ten nines) even below the critical time $T^*$, and does not show a plateau behavior. The behavior of the universal ansatz is consistent with the theoretical time lower bound of \cite{Hegerfeldt2013} for this Hamiltonian. However, this bound does not apply to the non-universal control ansatz. Instead, the minimal time required for the non-universal ansatz to implement a $Z$-rotation gate can be readily calculated from the Hamiltonian strength and is lower than the time limit for the universal ansatz. Therefore, in Fig. \ref{fig:T_NoBath}, the fidelity of the universal ansatz is very low when $T$ is less than the critical time $T^*$ but the fidelity of the non-universal ansatz is still high, even at times shorter than $T^*$. 
	
	After turning on the coupling to $n >0$ TLS, the fidelity is still above six nines and is weakly dependent on TLS count, as shown in Fig. \ref{fig:Non_uni_T}. Compared to the similar case for the universal ansatz in Fig. \ref{fig:TLS_single}, which is also plotted in  Fig. \ref{fig:Non_uni_T}, with highest fidelity around four nines, the fidelity of non-universal control ansatz is still higher. After tuning on Lindblad $T_1$ decay (purple points in Fig. \ref{fig:Non_uni_T}), the fidelity drops to around two nines, which is now similar to the performance of universal ansatz in Fig. \ref{fig:TLS_Lind}. Lindblad results for $n=0,1$ (not shown) are virtually identical to the result shown here for $n=2$, indicating that the fidelity dependence on the number of TLS $n$ is even weaker in the presence of $T_1$ decay.
	
    In the case of the Hadamard gate target (not shown), the straight fidelity without $T_1$ decay is about 0.5 for any number $n$ of TLS. After turning on the $T_1$ decay, the straight fidelity drops with the protocol duration $T$ but generally stays above 0.4. The straight fidelity for the same protocol duration with different numbers of TLS is almost the same and is well below 0.9 (i.e. a MLI less than 1).
	
	In general, the non-universal control shows better performance when applied to $Z$-rotation gates like the T gate. For these gates, the fidelity obtained with the non-universal control is higher than can be obtained with the universal control ansatz and the critical times are also shorter. However, when applied to other gates such as the Hadamard gate, the non-universal control ansatz is not effective and gives fidelities well below 0.9. We note that the fidelity achieved for the T gate here is higher than that obtained for quantum gates on a 4-level qudit similarly experiencing spontaneous decay while also undergoing dipolar interaction with $n$ TLS spins \cite{Reich_2015}.
    \subsection{Two Qubits} \label{sec:2qb}
    So far we have shown the results of bipartite Hamiltonian switching control on a single qubit. In this section we present the results of applying the method to two system qubits to implement the entangling CNOT gate. The two qubit model studied here is two superconducting qubits coupled to a primary bath of TLS, as described by Eq. \ref{eq:dp2qb}, with or without a secondary Lindblad bath. The control ansatz is given in Eq. \ref{eq:2qb_ansatz}. 
    
    \begin{figure}[h!]
	 	\centering
	 	\begin{subfigure}[b]{\figwidthHalf}
	 	    \includegraphics[width=\linewidth]{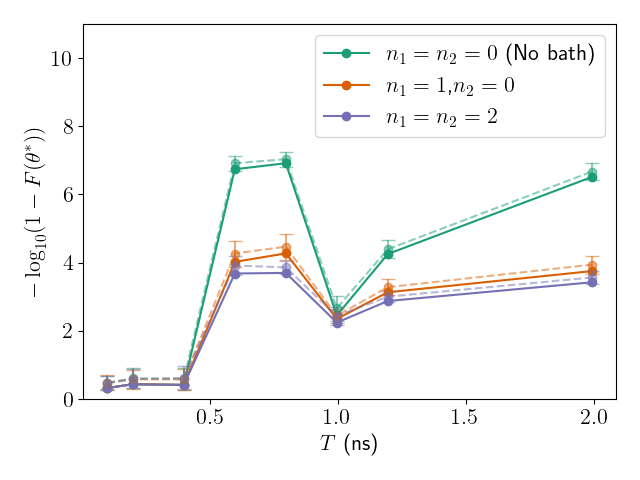}
	 	    \caption{2 qubit, $p=30$}
	 	    \label{fig:TLS_2qb_T}
	 	\end{subfigure}
	 	\begin{subfigure}[b]{\figwidthHalf}
	 	    \includegraphics[width=\linewidth]{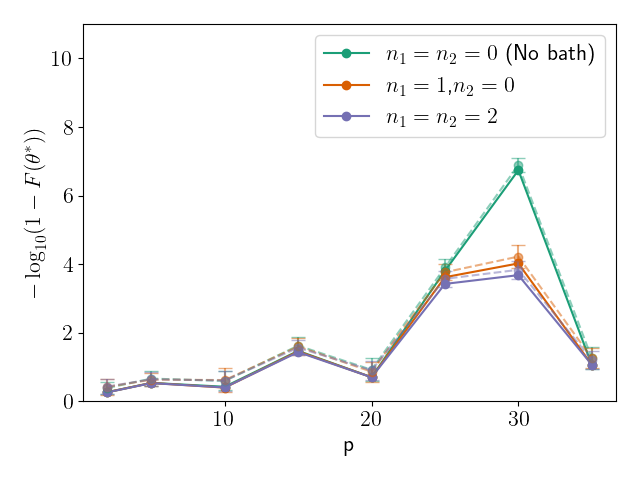}
	 	    \caption{2 qubit, $T=\frac{30}{16\pi}$ ns$\approx 0.6$ ns}
	 	    \label{fig:TLS_2qb_p}
	 	\end{subfigure}
	 	\caption{CNOT gate fidelity dependence on (a) total evolution time and (b) control depth for 2-qubit system coupled to $n$ TLS and optimized with PG. The coupling constants are unequal and evenly distributed in the relevant strength of superconducting devices. Each point is the best result of 5 parallel simulations with different initialization. For each pair of points with the same color and T value, the solid line is the actual optimization result and the pale line with error bar is the average state fidelity over $M=100$ Haar-random states. The fidelity, Eq. \ref{eq:AUfid}, is plotted as MLI, Eq. \ref{eq:MLI}. }
 	\label{fig:TLS_2qb}
	\end{figure}
	
	The results of 2-qubit control with only the primary TLS bath are plotted in Fig. \ref{fig:TLS_2qb}. Results for the case with no TLS are also plotted for comparison. The 2-qubit gate fidelity in the absence of TLS can reach seven nines, which shows that the ansatz is effective. After turning on the couplings, the fidelity drops but the dependence on total time and protocol depth has the same pattern as that with no couplings. The fidelity generally increases to about four nines after the critical time and critical depth, and is lower for smaller values of $T$ and $p$. The state fidelity of each protocol is also plotted in Fig. \ref{fig:TLS_2qb} (points with error bars, connected by pale dashed lines), showing good agreement with the optimization results. \\

    \begin{figure}[h!]
	 	\centering
	 	\begin{subfigure}[b]{\figwidthHalf}
	 	    \includegraphics[width=\linewidth]{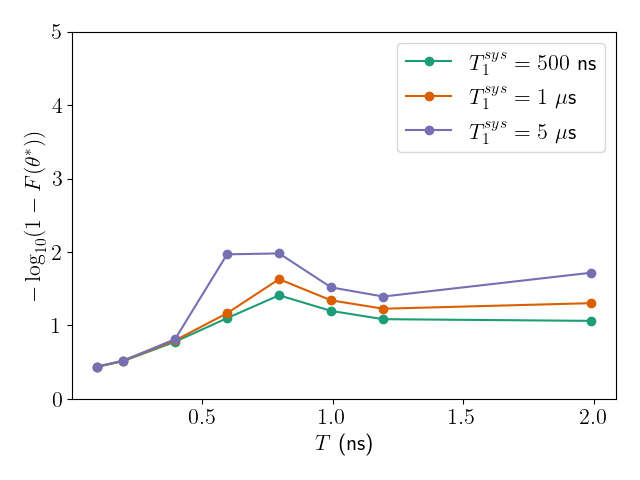}
	 	    \caption{$n_1= n_2= 0$ (No TLS)}
	 	    \label{fig:LindTLS_2qb_n0n0}
	 	\end{subfigure}
	 	\begin{subfigure}[b]{\figwidthHalf}
	 	    \includegraphics[width=\linewidth]{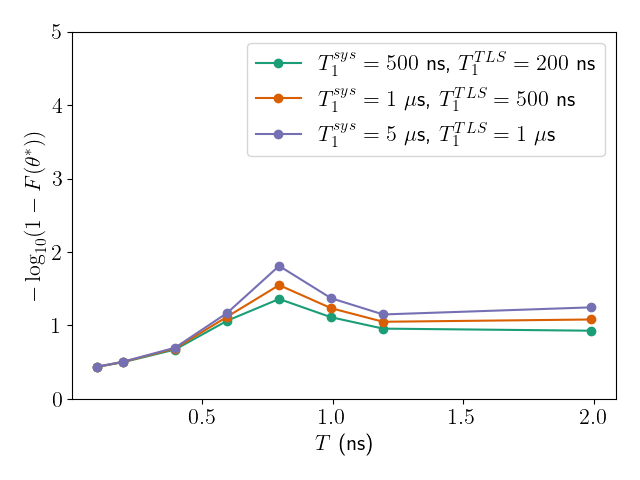}
	 	    \caption{$n_1=1, n_2=0$}
	 	    \label{fig:LindTLS_2qb_n1n0}
	 	\end{subfigure}
	 	\caption{CNOT gate fidelity dependence on total evolution time for 2-qubit system coupled to $n$ TLS under the effect of both TLS couplings and Lindblad decoherence optimized with PG. Each point is the best result of 3 parallel simulations with different initialization. The control depth is $p=30$. The fidelity, Eq. \ref{eq:ref_state_fid}, is plotted as MLI, Eq. \ref{eq:MLI}.}
 		\label{fig:LindTLS_2qb}
	\end{figure}
	
	The results of 2-qubit control with both a primary TLS bath and a secondary Lindblad bath are plotted in Fig. \ref{fig:LindTLS_2qb}. As expected, the fidelity is lowered compared to that without Lindblad decoherence in Fig. \ref{fig:LindTLS_2qb}. In general, the effect of the Lindblad bath in decreasing the gate fidelity is seen to be more significant than the effect of the TLS bath. In addition, given a fixed level of Lindblad noise, the fidelity with different numbers of TLS is very similar. The lowest MLI under the strongest Lindblad noise ($T_1^{sys} = 500 ns, T_1^{TLS} = 200 ns$) and one TLS is about 1.36. The critical time is not significantly affected by either the Lindblad noise strength or the coupling strength to the TLS. 
	
	Comparing Fig. \ref{fig:TLS_single} and Fig. \ref{fig:TLS_2qb}, we can see that the two-qubit gate fidelities are similar to one-qubit when the qubits are coupled only to the primary TLS bath. But when the qubits are also coupled to the secondary Lindblad bath, the fidelity of two-qubit control is lower than the one-qubit case. It is apparent that the Hamiltonian switching control of a two-qubit system is less robust to Lindblad noise. In both cases (i.e. with or without $T_1$ decay), the critical time $T^*$ and depth $p^*$ of two-qubit control are both larger than  one-qubit control.
 
    \section{Discussion}\label{sec:discussion}
    Table \ref{tab:summary} summarizes all the simulations carried out in this work using Hamiltonian switching.
    We focus our discussion of the results primarily on the single-qubit dipole-dipole coupling case studied in Sec. \ref{sec:Transmon} since this is the most relevant model for real devices. 
     
    The fidelity at the plateau region for all numbers of TLS $n=1-5$ is around four nines, lower than the six nines achieved for most of the test cases of the isotropic model, for which the coupling strength between the system qubits and bath spins is about three degrees of magnitudes stronger. Beyond the critical depth $p^*$, further increasing the protocol depth $p$ does not appear to improve the fidelity. Adjusting other protocol parameters is also found to be ineffective in increasing the fidelity level. To find the factors affecting the highest achievable fidelity level, we have performed additional simulations shown in Appendix \ref{sec:fid_factors}. These results show that performing the control for the system in a rotating frame with strong system-bath couplings achieve the highest fidelity level, which is then usually six to eight nines. Changing to either the lab frame or the weak coupling regime is found to lower the fidelity to around four nines. Systems for which control is carried out in both the lab frame and the weak coupling regime also have a fidelity around four nines. This was observed for both types of couplings to the primary bath of spins, i.e., isotropic and dipolar.   
     \begin{table}[h]
        \centering
        \begin{tabular}{c|c|c|c|c|c}
           Primary bath & Coupling strength & Ansatz & \# of qubits & Secondary bath  &  Highest MLI  \\
           \hline
           \begin{tabular}{@{}c@{}}Isotropic coupling,\\ rotating frame \end{tabular}& $\sim 1$ & universal & 1 & No& 8.14 \\ \hline
           \multirow{6}{*}{\begin{tabular}{@{}c@{}}Dipole-dipole coupling,\\ lab frame \end{tabular}} & \multirow{6}{*}{$5*10^{-4} - 5*10^{-3}$ } & \multirow{4}{*}{universal} & \multirow{2}{*}{1} & No & 5.15\\ \cline{5-6}
            &&&& Lindblad & 2.79 \\ \cline{4-6}
            &&&\multirow{2}{*}{2}& No & 4.27 (9.48)\footnote{Highest MLI obtained with secondary GRAPE optimization.}\\ \cline{5-6}
            &&&& Lindblad & 1.81 \\ \cline{3-6}
            && \multirow{2}{*}{non-universal} & \multirow{2}{*}{1} & No & 7.18\footnote{\label{footnote:table_non_uni}Only applicable to $Z$-rotation gates.}\\ \cline{5-6}
            &&&& Lindblad & 2.48\textsuperscript{\ref{footnote:table_non_uni}} \\ \hline
        \end{tabular}
        \caption{Summary of simulations carried out in this work}
        \label{tab:summary}
    \end{table}   
    
    To analyze these observations, we first computed the gradient and Hessian of several simulations, as described in Appendix. \ref{sec:fid_factors}. The largest components of the gradients for all four selected simulations were found to be approximately $10^{-3}$, which are relatively small, indicating proximity to critical points of the control landscape. The majority of the eigenvalues of the Hessian matrices were negative, indicative of saddle points that are nearly local maxima. Furthermore, in the weak coupling scenario, most of the eigenvalues had small magnitudes, suggesting a possible robustness of the control \cite{KosutBR:2022}.
    
   We also conducted a controllability analysis in Appendices \ref{sec:controllability_iso} and \ref{sec:controllability_dp} for the case of 2 bath spins, to understand whether the differences in precision achieved with Hamiltonian switching for the two types of bath coupling could be assigned to different extent of controllability.  The results are summarized in Table \ref{tab:controllability}.
    \begin{table}[h]
        \centering
        \begin{tabular}{c|c|c|c|c}
            Coupling type & Frame & Equal coupling strength? & Qubit controllable? & Spin bath controllable? \\
            \hline
            \multirow{4}{*}{Isotropic coupling} & \multirow{2}{*}{Rotating } & N & Y & Y \\ \cline{3-5}
            &&Y &Y&N\\ \cline{2-5}
            & \multirow{2}{*}{Lab } & N & Y & Y \\ \cline{3-5}
            &&Y &Y&Y\\ \hline
            \multirow{4}{*}{Dipole-dipole coupling} & \multirow{2}{*}{Rotating } & N & Y & N \\ \cline{3-5}
            &&Y &N&N\\ \cline{2-5}
            & \multirow{2}{*}{Lab } & N & Y & N \\ \cline{3-5}
            &&Y &Y&N\\ \hline
        \end{tabular}
        \caption{Summary of controllability of systems with 1 qubit and 2 bath spins. See  Appendices \ref{sec:controllability_iso} and \ref{sec:controllability_dp} for full details.}
        \label{tab:controllability}
    \end{table}
    
It is evident that the controllablity analysis is not able to explain the fidelity differences between rotating and lab frames, nor of course, differences due to the change of coupling strength. The controllability of the qubit and bath spins does not necessarily depend on whether they are described in the rotating or the lab frame: it depends only on whether the coupling strengths are all equal or not, rather than on the strength of the couplings relative to the system energy. Instead, the controllability affects only the critical depth $p^*$, as shown in Sec.\ref{sec:single_iso}.

We explored several options to improve the fidelity for the dipolar coupling case. One is to add an additional $\sigma^y$ control term, which ensures controllability of the system qubit regardless of the type and relative strength of the couplings to the bath spins. These calculations require extension of the two-Hamiltonian switching protocol to a four-Hamiltonian protocol and are summarized in Appendix. \ref{sec:4Ham}. The resulting simulation results show that the fidelity in the lab frame and strong coupling regime is now improved above six nines, but the fidelity in the other test cases is not significantly improved.

A more successful strategy derives from noting that for the specific choice of Hamiltonians $H_A$ and $H_B$ used in the switching Hamiltonian protocol in this work, Eq. (12), the effective control Hamiltonian reduces to alternating between $-2E\sigma_x$ and $+2E\sigma_x$, i.e., to a bang-bang form of control.  This suggests that we may further refine the strength of the effective control pulses obtained from the Hamiltonian switching method to allow variable strength of the $\sigma_x$ term during each variable time interval determined by optimization of the Hamiltonian switching protocol.  We carried this out by applying the GRAPE method as described in Sec. IV.B. The results for the dipole-dipole coupled systems in the lab frame and in the weak coupling regime were presented in Sec. \ref{sec:Transmon}, where we saw that the fidelity is now very significantly improved by this secondary optimization of the pulse strength. Note that this secondary optimization may also be understood as a generalization of the Hamiltonian switching protocol to allow variation of an additional parameter in each instance of $H_A$ and $H_B$, beyond just the duration of each instance.

 In general, our results indicate that the fidelities that can be achieved by the basic form of Hamiltonian switching given in Eq. (12) are primarily affected by the presence of bath $Z$ terms and by the relative strength of the qubit system-bath spin couplings and the system energies. The fact that the fidelities are lower in the weak coupling regime indicates that it is hard for the symmetric ('bang-bang') type of switching to cancel the effect of weak qubit-spin couplings, allowing the fidelities to reach only four or five nines.  However, the additional flexibility offered by a secondary optimization of the effective control strength from pulse to pulse using the GRAPE method allows this to be further improved to nine nines.

    \section{Summary and Conclusions}\label{sec:conclusions}
    In this work, we have developed a bipartite Hamiltonian switching control protocol for qubits coupled to a primary bath of two level systems and a Lindblad bath describing an extended environment. We applied this approach to systems of one and two qubits coupled to a primary bath of spins/TLS via either an isotropic interaction or a dipole-dipole interaction, and analysed the control of the system in this setting, as well as with a secondary Lindblad bath. The simulations for systems coupled only to the primary bath (spin or TLS bath) are optimized using a unitary gate fidelity measure, while the simulations for systems coupled to both a primary bath and a secondary Lindblad bath are optimized using a reference state fidelity measure.  The Hamiltonian control optimizes the time duration for application of each Hamiltonian, with constant amplitude in all time intervals.  In cases where the fidelity obtained with this basic Hamiltonian switching control was too low, we boosted the fidelity using a secondary optimization with GRAPE that also allows the amplitude of the Hamiltonian controls to be adjusted for each time interval.
	
	Our numerical results demonstrated the effectiveness of bipartite Hamiltonian switching control. In the presence of coupling to the primary bath, the fidelity of direct implementation of gates would drop below 0.9. However, with Hamiltonian switching, the fidelity in most of the test cases were improved to between $1-10^{-4}$ and $1-10^{-8}$.  Further refinement with a secondary GRAPE optimization brought the fidelity in all cases to $1-10^{-8}$ or higher, which is well within the thresholds of most contemporary quantum error correction codes. Applying the resulting protocols to specific states gives state fidelities of similar magnitude to the corresponding gate fidelities. 
	
	Analysis of the change of fidelity over the protocol evolution time reveals a general pattern that the fidelity first increases with time, then reaches a plateau region that is approximately constant as the protocol time increases further. A similar trend holds for the fidelity change as the control depth $p$, i.e., the number of Hamiltonian switching intervals, is increased. However, only with sufficient control depth $p$ can the behavior of an increase to a plateau value over total evolution time be observed, and the plateau value does not change significantly when the control depth is further increased. We refer to the turning points between the region of increasing fidelity and the subsequent plateau region as the critical time $T^*$ and critical depth $p^*$, for the dependence on protocol time $T$ and depth $p$ respectively.
	
	For qubit systems strongly coupled to the primary bath of spins with an isotropic interaction and a $Z$ gate target, the critical depth and critical time both increase with the number of bath spins. This is consistent with the simulated reduced dynamics in the absence of control, where different numbers of bath spins are known to give rather different system dynamics. However, for fixed system parameters, different target gates give similar fidelity values and almost identical patterns of change with control depth and evolution time. The most efficient protocol for the implementation of a gate is obtained using the critical time $T^*$ and critical depth $p^*$. Changing the coupling strengths between the system qubits and bath spins from all equal to variable strength was found to have a significant impact on the fidelity, with the critical depth $p^*$ being much larger for the variable coupling strength case. In most of the simulations with variable coupling strength, despite increased control depth, the highest fidelity reached in the isotropic model is significantly lower than that found in the equal-coupling case. This validates the claim in \cite{Arenz2014} that the variable coupling case is harder to control.
	
	In the case of qubits coupled to TLS through dipole-dipole interactions, with parameters emulating typical superconducting qubit systems, the coupling strengths are 2 to 3 magnitudes weaker than those we employed for the isotropic central spin model with variable values. With a $Z$ gate target, simulations with different TLS count $n$, control depth $p$ and optimization method were found to yield almost identical highest fidelity values, as well as identical dependence on the total protocol evolution time $T$. The critical time and depth are lowered to $T^* = \frac{4}{16\pi} \, \text{ns}\approx 0.08$ ns (4 time units) and $p^* = 5$, suggesting significantly easier implementation in experimental systems. In contrast to the isotropic central spin model, these values are independent of the number of TLS.  This is consistent with the weaker coupling and also with the simulated reduced dynamics, since the reduced dynamics of a qubit with no control fields but coupled by dipolar interactions to different numbers of TLS are almost the same up to 2 ns, a considerably longer time than all protocols in this work. The critical time and critical  depth are similar for Hadamard and T gate targets. The highest fidelity reached by Hamiltonian switching alone for qubits that are dipolar coupled to a bath of TLS is around $1-10^{-4}$, but was shown to be improved to  $1-10^{-9}$ by a secondary optimization by GRAPE that modulates the amplitudes of the switching Hamiltonians in addition to their duration. Introducing additional couplings to a secondary Lindblad bath causing $T_1$ decay of the system qubits lowers the fidelity, as expected. Nevertheless, a fidelity of approximately $1-10^{-3}$ on a single qubit can still be reached when the $T_1$ time of the qubit is 500 ns.
	
	The non-universal ansatz is not capable of implementing arbitrary single-qubit gates. While it can reach higher fidelities when implementing Z-rotations including the T gate, also with lower critical time than the universal ansatz, the fidelity is however very low when implementing other gates. Our analysis shows that while coupling to TLS changes the Lie algebraic structure of the coupled system-primary bath, it does not increase the fidelity per se when implementing either $Z$-rotation or Hadamard gates. In general, this non-universal ansatz is better suited for implementing $Z$-rotation gates like the T gate, but is not as capable for other gates like the Hadamard gate.
	
     This work demonstrated that bipartite Hamiltonian switching control for high fidelity quantum gates can very effectively mitigate environmental noise, particularly when the optimization of the switching Hamiltonian pulse durations is supplemented by secondary optimization of the switching Hamiltonian amplitudes. Our analysis revealed several interesting features of this method. 
     One notable feature is that with  a universal ansatz, the control properties depend only weakly on the target gate, suggesting  a generic robustness of the optimal bipartite control.
     It will be interesting in the future to further explore the robustness of this bipartite control scheme in a more systematic manner, e.g., examining its performance under diverse uncertainties using the recently developed classical averaging method~\cite{KosutBR:2022}. 
    For example, one can ask whether there are classes of imperfections for which control of switching time intervals and the fine-tuning their amplitudes is naturally more robust than standard pulse shaping. Many robustness issues are related to the topology of the quantum control landscape, i.e., the fidelity as a function of the control variables. In particular, at the top of the landscape where the fidelity is near one and the gradient is near zero, there is a large null space which allows for considerable design freedom, e.g., for a significant robustness in the choice of control parameters. An interesting question here is to compare the null space for Hamiltonian switching with that for standard pulse control. For example, if the key variable is the area under the pulse, for the standard case this is simply amplitude times a fixed time interval, whereas for Hamiltonian switching it is a fixed amplitude times a varying time-interval. The robustness properties will also depend on the accuracy of setting the pulse intervals and/or the pulse amplitudes. In some cases larger pulse amplitudes induce other phenomena, e.g., if too much energy is pumped into the system. In such situations, Hamiltonian switching by intrinsically fixing the pulse amplitudes may avoid disruptions or unwanted side effects. In general, we conclude that Hamiltonian switching bipartite control is a promising practical candidate for the implementation of high-fidelity quantum gates on NISQ devices.

	\section{Acknowledgements}
	ZY thanks Yulong Dong, Jiahao Yao, Lin Lin, Zhang Jiang, Zengzhao Li, Liwen Ko, Haoran Liao and Ian Convy for helpful discussions. This material is based on work supported by the U.S. Department of Energy, Office of Science, National Quantum Information Science Research Centers, Quantum Systems Accelerator. This work was also partially supported by a Google Quantum Research Award (Z. Y.), We also thank Berkeley Research Computing (BRC) group for providing time on their Savio computation cluster. RLK acknowledges partial support from the U.S. Department of Energy (DOE) under STTR Contract No. DESC0020618.

    \bibliography{references} 
    \pagebreak
    
    \appendix
    \section{Universality of 2-qubit Ansatz} \label{sec:2qb_uni_proof}
    The Lie algebra generated by the two-qubit ansatz Eq. \ref{eq:2qb_ansatz}, $\mathcal{L}_{\text{2qb}}$, without coupling to the bath is the same as that derived from 
    \begin{equation}
        \begin{aligned}
            H_Z =& \sigma^z_0+\omega \sigma^z_1+\gamma \sigma^z_0\sigma^z_1\\
            H_X =& \sigma^x_0+\sigma^x_1.
        \end{aligned}
    \end{equation}
    We note that we need to set $\omega\neq 1$ in order to make this ansatz universal, according to \cite{Morales2020}. From $[iH_Z,iH_X]$, we then get, up to a constant factor,
    \begin{equation}
        H_{\text{YZ}} = \sigma^y_0+\omega \sigma^y_1 +\gamma (\sigma^y_0\sigma^z_1+\sigma^z_0\sigma^y_1)
    \end{equation}
    From $[iH_{\text{YZ}},iH_X]$ and linear combination with $H_Z$, we get, up to a constant factor,
    \begin{equation}
        H^{(1)} = 2\sigma^y_0\sigma^y_1-\sigma^z_0\sigma^z_1.
    \end{equation}
    From $[iH^{(1)},iH_X]$ and linear combination with $H_{\text{YZ}}$, we get, up to a constant factor,
    \begin{equation}
        H_Y = \sigma^y_0+\omega \sigma^y_1
    \end{equation}
    From $[[iH_Y,iH_X],iH_Y]$, we get, up to a constant factor,
    \begin{equation}
        H_{X1} = \sigma^x_0+\omega^2\sigma^x_1
    \end{equation}
    By evaluating the commutator and taking linear combinations between $H_X, H_{X1}, H_Y$, we can readily show that
    \begin{equation}
        i\sigma^x_0, i\sigma^y_0, i\sigma^z_0,i\sigma^x_1, i\sigma^y_1, i\sigma^z_1 \in \mathcal{L}_{\text{2qb}},
    \end{equation}
    which means that arbitrary single-qubit rotations on either qubit can be generated. Additionally, arbitrary single-qubit unitaries combined with any entangling two-qubit operator can generate any SU(4) unitary \cite{Morales2020,Brylinski2001}. Furthermore, we can readily show that the entangling $\sigma^z_0\sigma^z_1$ operator is in the Lie algebra by subtracting the single-qubit $\sigma^z$ terms from $H_Z$. Therefore, this ansatz is universal when there are two qubits.
    \section{Factors Impacting Fidelity}\label{sec:fid_factors}
\begin{figure}[h!]
        \centering
        \begin{subfigure}[b]{\figwidthThird}
            \includegraphics[width=\linewidth]{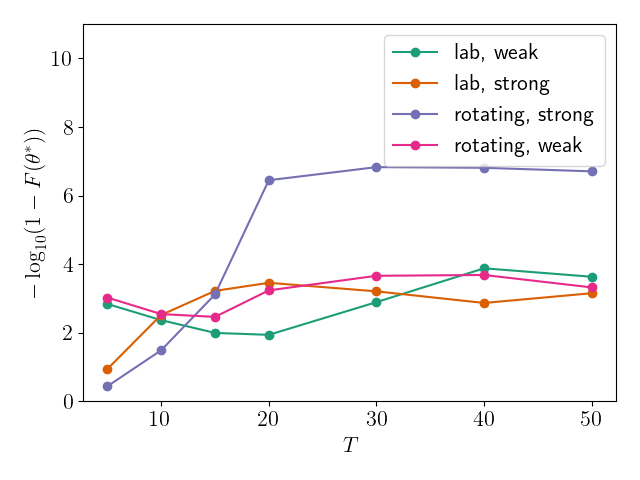}
            \caption{Equal isotropic couplings}
        \end{subfigure}
        \begin{subfigure}[b]{\figwidthThird}
            \includegraphics[width=\linewidth]{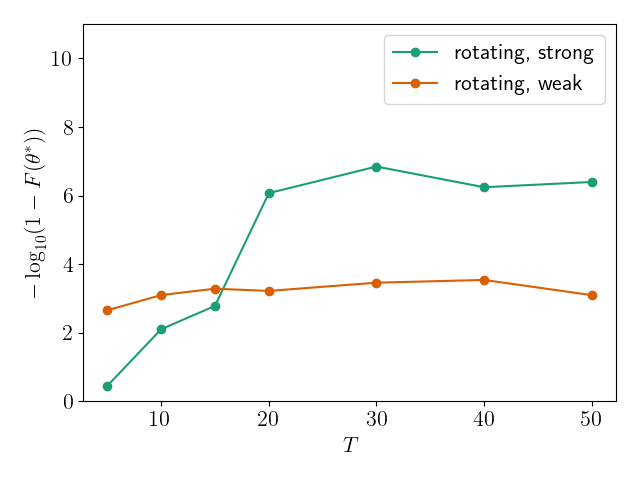}
            \caption{Variable isotropic couplings}
        \end{subfigure}
        \begin{subfigure}[b]{\figwidthThird}
            \includegraphics[width=\linewidth]{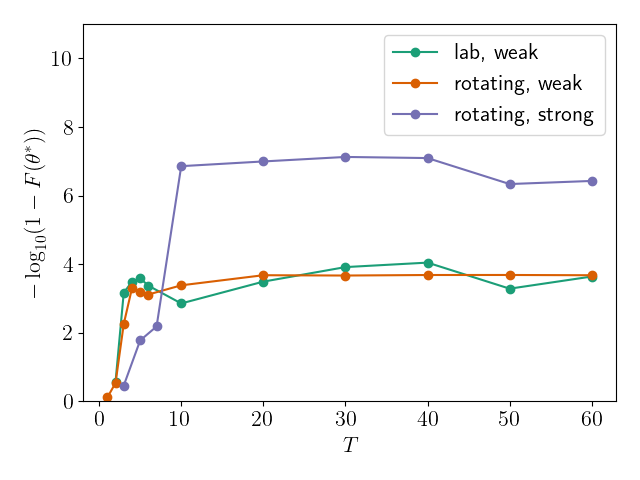}
            \caption{Equal dipolar couplings}
        \end{subfigure}
        \caption{Fidelity dependence on evolution time for (a) and (b) isotropically and (c) dipole-dipole coupled systems optimized with PG. The target gate is $Z$. The coupling constants are (a) and (c) equal and (b) unequal. The number of bath spins for all the plots is $n=2$. For each line, lab means lab frame (i.e. The Hamiltonian includes bath $\sigma^Z$ terms) and rotating means rotating frame (i.e. no bath $\sigma^z$ terms), and strong means strong coupling strength (i.e. in the same order of magnitude as the system energy splitting) and weak means weak coupling strength (i.e. about $10^{-3}$ relative to the system energy splitting). Each point is the best result of 3 or 5 parallel simulations with different initialization. The fidelity, Eq. \ref{eq:AUfid}, is plotted as MLI, Eq. \ref{eq:MLI}. }
        \label{fig:fid_factors}
    \end{figure}

    To find the system parameters that impact the highest fidelity can be achieved in the simulations, further single-qubit simulations for both types of CS couplings are done. We set $E=1$. In the strong coupling regime, we have $A_q = E=1$ when the couplings are equal and $A_q$ uniformly distributed in $[1,2]$ when they are variable. In the weak coupling regime for isotropic couplings, we have $A_q = E=0.005$ when the couplings are equal and $A_q$ uniformly distributed in $[5,10]*10^{-3}$ when they are variable. In the weak coupling regime for dipole-dipole couplings, we have $A_q = E=0.005$ when the couplings are equal and $A_q$ uniformly distributed in $[5,50]*10^{-4}$ when they are variable. For all the cases, we set $\Delta_q=0$ for rotating frame and $\Delta_q = 1.0,1.1$ for lab frame, as we focus on the case of $n=2$ bath spins/TLS. The results are shown in Fig. \ref{fig:fid_factors}.

    To further understand the numerical properties of the optimal protocols, we evaluated the gradient and Hessian of the simulation results with $T=30$ units in Fig. \ref{fig:fid_factors}a. The $l$-infinity norm (i.e. largest absolute value of the components) of the gradients and the eigenvalues of Hessian matrices are summarized in Table. \ref{tab:grad_hessian}
    \begin{table}[h]
        \centering
        \begin{tabular}{c|c|c}
           Model & Rotating Frame, Strong Couplings & Lab Frame, Strong Couplings\\
           \hline
           MLI & 6.83 & 3.21 \\
           \hline
           $\left|\nabla F(\theta)|_{\theta=\theta^*} \right|_{\infty}$&$1.48 \times 10^{-3}$&$1.46 \times 10^{-3}$ \\
           \hline
           Hessian Eigenvalues &
           \includegraphics[width=0.3\linewidth]{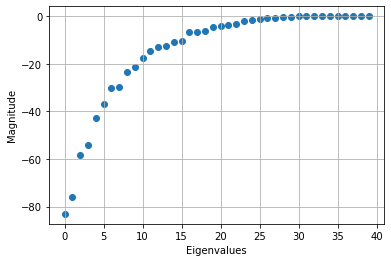}&
           \includegraphics[width=0.3\linewidth]{Plots/Supplementary/grad_heis/iso_rot_strong.png}\\
           \hline
           Model & Rotating Frame, Weak Couplings & Lab Frame, Weak Couplings\\
           \hline
           MLI & 3.66& 2.89 \\
           \hline
           $\left|\nabla F(\theta)|_{\theta=\theta^*} \right|_{\infty}$&$0.91 \times 10^{-3}$&$2.25 \times 10^{-3}$ \\
           \hline
           Hessian Eigenvalues &
           \includegraphics[width=0.3\linewidth]{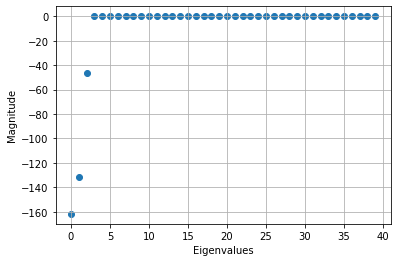}&
           \includegraphics[width=0.3\linewidth]{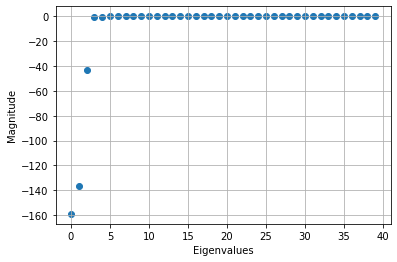}\\
           \hline
        \end{tabular}
        \caption{$l$-infinity norm (i.e. largest absolute value of the components) of the gradients and plots of the eigenvalues of Hessian matrices for selected simulations.}
        \label{tab:grad_hessian}
    \end{table}
    
    \section{Four-Hamiltonian Switching}\label{sec:4Ham}
    To improve the fidelity in the lab frame and weak coupling regime, we tested an ansatz containing an additional $\sigma^y$ term, which results in switching between the following four Hamiltonians:
    \begin{equation} \label{eq:TLS_ABCD}
	\begin{aligned}
	H_A^{\text{dipole}} = -\frac{E}{2} \sigma^z_0 + 2E \sigma_0^x + \frac{3}{2}E \sigma_0^y -\sum_{q=1}^n \frac{\Delta_q}{2}\sigma^z_q + \sum_{q=1}^{n} \frac{A_q}{2} (\sigma^+_0\sigma^-_q + \sigma^-_0\sigma^+_q )\\
	H_B^{\text{dipole}} = -\frac{E}{2} \sigma^z_0 + 2E \sigma_0^x - \frac{3}{2}E \sigma_0^y -\sum_{q=1}^n \frac{\Delta_q}{2}\sigma^z_q + \sum_{q=1}^{n} \frac{A_q}{2} (\sigma^+_0\sigma^-_q + \sigma^-_0\sigma^+_q ) \\
         H_C^{\text{dipole}} = -\frac{E}{2} \sigma^z_0 - 2E \sigma_0^x + \frac{3}{2}E \sigma_0^y -\sum_{q=1}^n \frac{\Delta_q}{2}\sigma^z_q + \sum_{q=1}^{n} \frac{A_q}{2} (\sigma^+_0\sigma^-_q + \sigma^-_0\sigma^+_q ) \\
         H_D^{\text{dipole}} = -\frac{E}{2} \sigma^z_0 - 2E \sigma_0^x - \frac{3}{2}E \sigma_0^y -\sum_{q=1}^n \frac{\Delta_q}{2}\sigma^z_q + \sum_{q=1}^{n} \frac{A_q}{2} (\sigma^+_0\sigma^-_q + \sigma^-_0\sigma^+_q ).
	\end{aligned}
    \end{equation}

\begin{figure}[h!]
        \centering
        \begin{subfigure}[b]{\figwidthThird}
            \includegraphics[width=\linewidth]{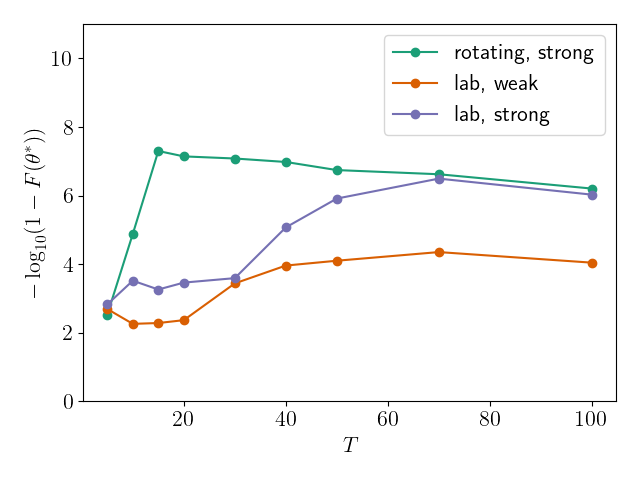}
            \caption{Equal isotropic couplings}
        \end{subfigure}
        \begin{subfigure}[b]{\figwidthThird}
            \includegraphics[width=\linewidth]{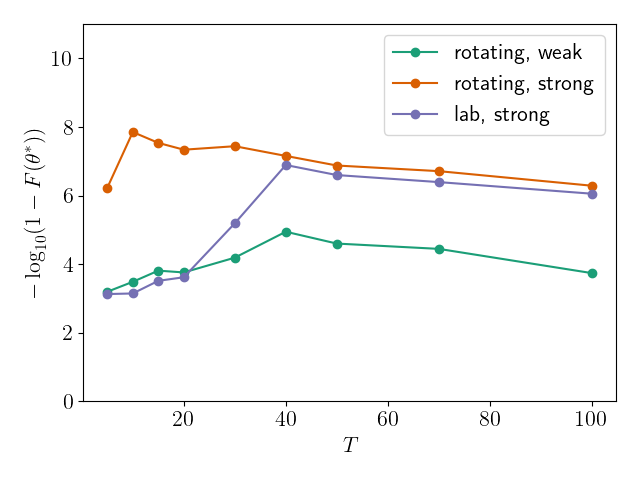}
            \caption{Equal dipolar couplings}
        \end{subfigure}
        \caption{Fidelity dependence on evolution time $T$ for (a) isotropically and (b) dipole-dipole coupled systems optimized with PG using 4-Hamiltonian switching. The target gate is $Z$. The coupling constants are equal. The number of bath spins for all the plots is $n=2$ and the control depth is $p=20$. For each line, ``lab" means lab frame (i.e., the Hamiltonian includes bath $\sigma^Z$ terms), ``rotating" means rotating frame (i.e., no bath $\sigma^z$ terms), ``strong" means strong coupling strength (i.e., of the same order of magnitude as the system energy splitting), and ``weak", means weak coupling strength (i.e. about $10^{-3}$ relative to the system energy splitting). Each point is the best result of 5 parallel simulations with different initialization. The fidelity, Eq. \ref{eq:AUfid}, is plotted as MLI, Eq. \ref{eq:MLI}. }
        \label{fig:4Ham}
    \end{figure}

    It can be readily shown that this ansatz is universal on the qubit. The fidelity dependence on evolution time $T$ is plotted in Fig. \ref{fig:4Ham}.
    \section{Controllability Analysis: Isotropic Couplings}\label{sec:controllability_iso}
    The analysis in this section focus on the case of $n=2$ bath spins. For each model in this section and Appendix. \ref{sec:controllability_dp}, we will explore the single and two-spin Pauli terms that are in the Lie algebra that are generated by the corresponding Hamiltonians. A spin is fully controllable if all three single-spin Pauli terms are in the Lie algebra, as the corresponding Lie group is equal to the special unitary group \cite{Arenz2014}. For one qubit coupled to two bath spins, all the single and two spin terms are 
    \begin{equation}\label{eq:Paulis}
        \begin{aligned}
        &i\sigma^x_{\alpha},i\sigma^y_{\alpha},i\sigma^z_{\alpha}, \\
        &i\sigma^x_{\alpha}\sigma^x_{\beta} ,i\sigma^x_{\alpha}\sigma^y_{\beta}, i\sigma^x_{\alpha}\sigma^z_{\beta},\\
        &i\sigma^y_{\alpha}\sigma^x_{\beta} ,i\sigma^y_{\alpha}\sigma^y_{\beta}, i\sigma^y_{\alpha}\sigma^z_{\beta}, \\
        &i\sigma^z_{\alpha}\sigma^x_{\beta} ,i\sigma^z_{\alpha}\sigma^y_{\beta}, i\sigma^z_{\alpha}\sigma^z_{\beta},    
        \end{aligned}
    \end{equation}
    where $\alpha, \beta \in \{0,1,2\}$.
    \subsection{Rotating Frame}
    According to the controllability analysis for isotropically coupled systems in \cite{Arenz2014}, when the coupling strengths are equal, the qubit is controllable and the CS bath is controllable in each subspace conserving total angular momentum, while when the coupling strengths are variable, the qubit and the full CS bath are both controllable, using a control ansatz with identical Lie algebra to that in this work.
    \subsection{Lab Frame}
    The Lie algebra generated by the two switching Hamiltonians is identical to that generated by
    \begin{equation}
        \begin{aligned}
        H^A_0 &= \sigma^z_0+g(\sigma^x_0\sigma^x_1+\sigma^y_0\sigma^y_1+\sigma^z_0\sigma^z_1)+h(\sigma^x_0\sigma^x_2+\sigma^y_0\sigma^y_2+\sigma^z_0\sigma^z_2)+\sigma^z_1+\epsilon \sigma^z_2\\
        H^A_c &= \sigma^x_0,
        \end{aligned}
    \end{equation}
    where $g,h$ are the coupling constants and can be equal or not, and the bath energy splittings are $1,\epsilon$ ($\epsilon \neq 1$). We define $A_1 = iH^A_0$ and $A_2 = iH^A_c$. With the commutator $[A_1, A_2]$, we have, up to a constant\begin{equation}
        A_3 = i(\sigma^y_0+g(\sigma^y_0\sigma^z_1 - \sigma^z_0\sigma^y_1) +h(\sigma^y_0\sigma^z_2-\sigma^z_0\sigma^y_2)).
    \end{equation}
    From $[A_3, A_2]$, we get, up to a constant
    \begin{equation}
        A_4 = i(\sigma^z_0+g(\sigma^y_0\sigma^y_1+\sigma^z_0\sigma^z_1)+h(\sigma^y_0\sigma^y_2+\sigma^z_0\sigma^z_2)).
    \end{equation}
    Then
    \begin{equation}
        A_5 = i(A_1-A_4) = i(g\sigma^x_0\sigma^x_1+h\sigma^x_0\sigma^x_2+\sigma^z_1+\epsilon \sigma^z_2).
    \end{equation}
    The nested commutator $[[A_3,A_4],A_5]$ yields up to a constant
    \begin{equation}
        A_6 = i(2g^2\sigma^y_1+h^2(1+\epsilon)\sigma^y_2+gh(h-g)\sigma^z_1\sigma^y_2+gh(g-h)\sigma^y_1\sigma^z_2+gh\sigma^x_0\sigma^x_1\sigma^y_2+gh\epsilon \sigma^x_0\sigma^y_1\sigma^x_2).
    \end{equation}
    We then define
    \begin{equation}\label{eq:iso_lab_rec_full}
        \begin{aligned}
        A^{(s)} &= i(k_1^{(s)}\sigma^y_1 + k_2^{(s)}\sigma^y_2 + k_3^{(s)}\sigma^z_1\sigma^y_2 + k_4^{(s)}\sigma^y_1\sigma^z_2 + k_5^{(s)}\sigma^x_0\sigma^x_1\sigma^y_2 + k_6^{(s)}\sigma^x_0\sigma^y_1\sigma^x_2), \\
        A^{(s)} &= [[A^{(s-1)},A_5],A_5]\\ 
        A^{(1)} &= A_6, \\
        \end{aligned}
    \end{equation}
    and by evaluating the nested commutator with $A_5$, i.e., $[[\cdot, A_5],A_5]$, we then get the recursive formulae
    \begin{equation}\label{eq:iso_lab_recursive}
        \begin{aligned}
        k_1^{(s)} & = -k_1^{(s-1)}(g^2+1)\\
        k_2^{(s)} & = -k_2^{(s-1)}(h^2+\epsilon^2)\\
        k_3^{(s)} & = -(h^2+g^2+\epsilon^2)k_3^{(s-1)} - 2ghk_4^{(s-1)} + gk_5^{(s-1)}+2g\epsilon k_6^{(s-1)}\\
        k_4^{(s)} & = -2ghk_3^{(s-1)} -(h^2+g^2+1)k_4^{(s-1)}  + 2hk_5^{(s-1)}+h\epsilon k_6^{(s-1)}\\
        k_5^{(s)} & = gk_3^{(s-1)} + 2hk_4^{(s-1)} - (1+\epsilon^2+h^2)k_5^{(s-1)} -2\epsilon k_6^{(s-1)}\\
        k_6^{(s)} & = 2g\epsilon k_3^{(s-1)}+h\epsilon k_4^{(s-1)}- 2\epsilon k_5^{(s-1)} - (1+\epsilon^2+g^2)k_6^{(s-1)}\\
        \end{aligned}
    \end{equation}    
    Inspired by the method in \cite{Arenz2014}, we try to isolate the single spin terms $i\sigma^y_1,i\sigma^y_2$ by taking linear combinations between $A^{(s)}$ formulae. If there exists six such coefficients, $s_1,s_2,s_3,s_4,s_5,s_6$, such that the determinant of the following matrix 
    \begin{equation} \label{eq:coef_mtx}
        \begin{pmatrix}
        k_1^{(s_1)} & k_2^{(s_1)} &k_3^{(s_1)} & k_4^{(s_1)}&k_5^{(s_1)} & k_6^{(s_1)}\\
        k_1^{(s_2)} & k_2^{(s_2)} &k_3^{(s_2)} & k_4^{(s_2)}&k_5^{(s_2)} & k_6^{(s_2)}\\
        k_1^{(s_3)} & k_2^{(s_3)} &k_3^{(s_3)} & k_4^{(s_3)}&k_5^{(s_3)} & k_6^{(s_3)}\\
        k_1^{(s_4)} & k_2^{(s_4)} &k_3^{(s_4)} & k_4^{(s_4)}&k_5^{(s_4)} & k_6^{(s_4)}\\
        k_1^{(s_5)} & k_2^{(s_5)} &k_3^{(s_5)} & k_4^{(s_5)}&k_5^{(s_5)} & k_6^{(s_5)}\\
        k_1^{(s_6)} & k_2^{(s_6)} &k_3^{(s_6)} & k_4^{(s_6)}&k_5^{(s_6)} & k_6^{(s_6)}\\
        \end{pmatrix}
    \end{equation}
     is non-zero, then one can get $i\sigma^y_1,i\sigma^y_2$ from the linear combination between $A^{(s_1)}, A^{(s_2)}, A^{(s_3)}, A^{(s_4)} ,A^{(s_5)}, A^{(s_6)}$. Numerical tests for $s_1,s_2,s_3,s_4,s_5,s_6 = 1,2,3,4,5,6$ show that the determinant of the matrix Eq. \ref{eq:coef_mtx} is generally non-zero for both $g=h$ and $g\neq h$. This implies that $i\sigma^y_1,i\sigma^y_2 \in \mathcal{L}$. Other single and two-spin terms in Eq. \ref{eq:Paulis} can be obtained by the following steps:
    \begin{itemize}
        \item from $[i\sigma^y_1,A_4]$, we get, up to a constant, $i\sigma^z_0\sigma^x_1$ and similarly $i\sigma^z_0\sigma^x_2$
        \item from $[i\sigma^y_1,i\sigma^z_0\sigma^x_1]$, we get, up to a constant,, $i\sigma^z_0\sigma^z_1$ and similarly $i\sigma^z_0\sigma^z_2$
        \item from $[i\sigma^x_0,i\sigma^z_0\sigma^x_1]$, we get up, to a constant $i\sigma^y_0\sigma^x_1$, and similarly $i\sigma^y_0\sigma^x_1,i\sigma^y_0\sigma^x_2,i\sigma^y_0\sigma^z_1,i\sigma^y_0\sigma^z_2$
        \item from $[[\sigma^y_1,A_5],i\sigma^z_0\sigma^z_1]$, we get up, to a constant, $i(g\sigma^y_0-\sigma^z_0\sigma^y_1)$. Similarly, we obtain $i(h\sigma^y_0-\epsilon \sigma^z_0\sigma^y_2)$. Taking linear combinations of these terms with $A_3, \sigma^y_0\sigma^z_1,\sigma^y_0\sigma^z_2$, we get,, up to a constant $i\sigma^y_0$
        \item other terms in Eq. \ref{eq:Paulis} can be readily generated
    \end{itemize}
    These steps hold for both $g=h$ and $g\neq h$. The Lie algebra generated in this way guarantees that the qubit and bath spins are all controllable as all the single and two-spin terms are in the Lie algebra. \cite{Burgarth2009,Arenz2014}.

    \section{Controllability Analysis: Dipole-dipole Couplings}\label{sec:controllability_dp}
    The analysis in this section focuses on the case of $n=2$ bath spins.
    \subsection{Rotating Frame}
    The Lie algebra generated by the two switching Hamiltonians is identical to that generated by
    \begin{equation}\label{eq:Lie_dp_rot_Ham}
    \begin{aligned}
    H_0^B &= \sigma^z_0+g(\sigma^x_0\sigma^x_1+\sigma^y_0\sigma^y_1)+h(\sigma^x_0\sigma^x_2+\sigma^y_0\sigma^y_2)\\
    H_c^B &= \sigma^x_0
    \end{aligned}
    \end{equation}
    where $g,h$ are the coupling constants and can be equal or not. Similarly, we define $B_1 = iH_0^B$ and $B_2 = iH_c^B$. With the commutator $[B_1, B_2]$, we have, up to a constant,
    \begin{equation}
        B_3 = i(\sigma^y_0-g\sigma^z_0\sigma^y_1 -h\sigma^z_0\sigma^y_2).
    \end{equation}
    From $[B_3, B_2]$, we get, up to a constant
    \begin{equation}
        B_4 = i(\sigma^z_0+g\sigma^y_0\sigma^y_1+h\sigma^y_0\sigma^y_2).
    \end{equation}
    Then
    \begin{equation}
        B_5 = i(B_1-B_4) = i(g\sigma^x_0\sigma^x_1+h\sigma^x_0\sigma^x_2).
    \end{equation}
    From $[[B_5,B_3],B_3]$, we get, up to a constant
    \begin{equation}
        B_6 = i(g^2\sigma^z_1+h^2\sigma^z_2+g(h^2+1)\sigma^x_0\sigma^x_1+h(g^2+1)\sigma^x_0\sigma^x_2).
    \end{equation}
     We can similarly define
     \begin{equation}
        \begin{aligned}
        B^{(s)} &= i(l_1^{(s)}\sigma^z_1 + l_2^{(s)}\sigma^z_2 + l_3^{(s)}\sigma^x_0\sigma^x_1+ l_4^{(s)}\sigma^x_0\sigma^x_2), \\
        B^{(s)} &= [[B^{(s-1)},B_3],B_3]\\ 
        B^{(1)} &= B_6, \\
        \end{aligned}
    \end{equation}
    and by evaluating the nested commutators with $B_3$, i.e., $[[\cdot, B_3],B_3]$, we then get the recursive formulae
    \begin{equation}\label{eq:recursive_dp_rot}
        \begin{aligned}
        l_1^{(s)} & =  l_1^{(s-1)}g^2 +  l_3^{(s-1)}g\\
        l_2^{(s)} & = l_2^{(s-1)}h^2 + l_4^{(s-1)}h\\
        l_3^{(s)} & = l_1^{(s-1)}g+l_3^{(s-1)}(h^2+1)\\
        l_4^{(s)} & = l_2^{(s-1)}h+l_4^{(s-1)}(g^2+1).
        \end{aligned}
    \end{equation}
    \subsubsection{Variable Couplings}
   Inspired by the method in \cite{Arenz2014}, we try to isolate the single spin terms $i\sigma^z_1,i\sigma^z_2$ by taking linear combinations between $B^{(s)}$ formulae. If there exists four such coefficients, $s_1,s_2,s_3,s_4$, such that the determinant of the following matrix 
    \begin{equation} \label{eq:coef_mtx_dp_rot}
        \begin{pmatrix}
        l_1^{(s_1)} & l_2^{(s_1)} &l_3^{(s_1)} & l_4^{(s_1)}\\
        l_1^{(s_2)} & l_2^{(s_2)} &l_3^{(s_2)} & l_4^{(s_2)}\\
        l_1^{(s_3)} & l_2^{(s_3)} &l_3^{(s_3)} & l_4^{(s_3)}\\
        l_1^{(s_4)} & l_2^{(s_4)} &l_3^{(s_4)} & l_4^{(s_4)}
        \end{pmatrix}
    \end{equation}
     is non-zero, then one can get $i\sigma^z_1,i\sigma^z_2$ from the linear combination between $B^{(s_1)}, B^{(s_2)}, B^{(s_3)}, B^{(s_4)}$. Numerical tests for $s_1,s_2,s_3,s_4 = 1,2,3,4$ show that the determinant of the matrix Eq. \ref{eq:coef_mtx_dp_rot} is generally non-zero for $g\neq h$. This implies $i\sigma^z_1,i\sigma^z_2 \in \mathcal{L}$. With commutators and linear combinations between $i\sigma^z_1,i\sigma^z_2$ and $B_3,B_4,B_5$, the following single and two-qubit/spin terms can be generated.
    \begin{equation}
        i\sigma_0^{\alpha}, i\sigma^z_a, i\sigma_0^{\alpha}\sigma_a^{\beta}
    \end{equation}
    Where $\alpha =x,y,z$, $\beta =x,y$ and $a=1,2$. The qubit is controllable but the full system is not.
    \subsubsection{Equal Couplings}
    When $g=h$, the determinant \ref{eq:coef_mtx_dp_rot} is always zero. We then attempted to get $i\sigma^z_1,i\sigma^z_2$ from other commutators but were not successful. Therefore, the only single spin term that we have found to be in the Lie algebra generated by the Hamiltonians Eq. \ref{eq:Lie_dp_rot_Ham} is $\sigma^x$ on the qubit, and thus it appears that neither the qubit nor the bath spins are controllable.

    \subsection{Lab Frame}
    The Lie algebra generated by the two switching Hamiltonians is identical to that generated by
        \begin{equation}
        \begin{aligned}
        H_0^C &= \sigma^z_0+g(\sigma^x_0\sigma^x_1+\sigma^y_0\sigma^y_1)+h(\sigma^x_0\sigma^x_2+\sigma^y_0\sigma^y_2)+\sigma^z_1+\epsilon \sigma^z_2\\
        H_c^C &= \sigma^x_0,
        \end{aligned}
    \end{equation}
    where $g,h$ are the coupling constants and can be equal or not, and the bath energy splittings are $1,\epsilon$ ($\epsilon \neq 1$). We define $C_1 = iH^C_0$ and $C_2 = iH^C_c$. With the commutator $[C_1, C_2]$, we have, up to a constant,
    \begin{equation}
        C_3 = i(\sigma^y_0-g\sigma^z_0\sigma^y_1 -h\sigma^z_0\sigma^y_2).
    \end{equation}
    From $[C_3, C_2]$, we get, up to a constant,
    \begin{equation}
        C_4 = i(\sigma^z_0+g\sigma^y_0\sigma^y_1+h\sigma^y_0\sigma^y_2).
    \end{equation}
    Then
    \begin{equation}
        C_5 = i(C_1-C_4) = i(g\sigma^x_0\sigma^x_1+h\sigma^x_0\sigma^x_2+\sigma^z_1+\epsilon \sigma^z_2)
    \end{equation}
    From $[[C_5, C_3],C_3]$, we get, up to a constant,
    \begin{equation}
        C_6 = i(2g^2\sigma^z_1+h^2(1+\epsilon)\sigma^z_2+g(h^2+2)\sigma^x_0\sigma^x_1+h(g^2+1+\epsilon)\sigma^x_0\sigma^x_2).
    \end{equation}
    We can similarly define
     \begin{equation}
        \begin{aligned}
        C^{(s)} &= i(m_1^{(s)}\sigma^z_1 + m_2^{(s)}\sigma^z_2 + m_3^{(s)}\sigma^x_0\sigma^x_1+ m_4^{(s)}\sigma^x_0\sigma^x_2), \\
        C^{(s)} &= [[C^{(s-1)},C_3],C_3]\\ 
        C^{(1)} &= C_6, \\
        \end{aligned}
    \end{equation}
    and by evaluation of the nested commutators with $C_3$, i.e., $[[\cdot, C_3],C_3]$, we get the recursive formulae
    \begin{equation}\label{eq:recursive_dp_lab}
        \begin{aligned}
        m_1^{(s)} & =  m_1^{(s-1)}g^2 +  m_3^{(s-1)}g\\
        m_2^{(s)} & = m_2^{(s-1)}h^2 + m_4^{(s-1)}h\\
        m_3^{(s)} & = m_1^{(s-1)}g+m_3^{(s-1)}(h^2+1)\\
        m_4^{(s)} & = m_2^{(s-1)}h+m_4^{(s-1)}(g^2+1).
        \end{aligned}
    \end{equation}
    Inspired by the method in \cite{Arenz2014}, we try to isolate the single spin terms $i\sigma^z_1,i\sigma^z_2$ by taking linear combinations between $C^{(s)}$ formulae. If there exists six such coefficients, $s_1,s_2,s_3,s_4$, such that the determinant of the following matrix
    \begin{equation} \label{eq:coef_mtx_dp_lab}
        \begin{pmatrix}
        m_1^{(s_1)} & m_2^{(s_1)} &m_3^{(s_1)} & m_4^{(s_1)}\\
        m_1^{(s_2)} & m_2^{(s_2)} &m_3^{(s_2)} & m_4^{(s_2)}\\
        m_1^{(s_3)} & m_2^{(s_3)} &m_3^{(s_3)} & m_4^{(s_3)}\\
        m_1^{(s_4)} & m_2^{(s_4)} &m_3^{(s_4)} & m_4^{(s_4)}
        \end{pmatrix}
    \end{equation}
     is non-zero, then one can get $i\sigma^z_1,i\sigma^z_2$ from the linear combination between $C^{(s_1)}, C^{(s_2)}, C^{(s_3)}, C^{(s_4)}$. Numerical tests for $s_1,s_2,s_3,s_4 = 1,2,3,4$ show that the determinant of the matrix Eq. \ref{eq:coef_mtx_dp_lab} is generally non-zero for both $g=h$ and $g\neq h$. This implies $i\sigma^z_1,i\sigma^z_2 \in \mathcal{L}$. With commutators and linear combinations between $i\sigma^z_1,i\sigma^z_2$ and $C_3,C_4,C_5$, the following single and two-qubit/spin terms can be generated.
    \begin{equation}
        i\sigma_0^{\alpha}, i\sigma^z_a, i\sigma_0^{\alpha}\sigma_a^{\beta}
    \end{equation}
    Where $\alpha =x,y,z$, $\beta =x,y$ and $a=1,2$. These steps hold for both $g=h$ and $g\neq h$. The qubit is controllable but the full system is not.
    
\end{document}